\documentclass[a4paper,11pt,fleqn]{article}
\usepackage{jheppub}

\usepackage[utf8]{inputenc}
\usepackage{xcolor}
\usepackage{relsize}
\usepackage{graphics}
\usepackage{epstopdf}
\usepackage{hyperref}
\usepackage{mathrsfs}
\usepackage{ragged2e}
\usepackage{amssymb}
\usepackage{placeins}

\usepackage{comment}
\usepackage{float}
\usepackage[normalem]{ ulem }
\usepackage{amsthm}
\usepackage{amsmath}
\usepackage{nicefrac}
\usepackage{mathtools}
\usepackage{subcaption}
\usepackage[noabbrev]{cleveref}
\usepackage[nolist]{acronym}
\usepackage{cancel}
\usepackage[textsize=tiny,backgroundcolor=yellow]{todonotes}

\def\ie{{\it i.e.},~}
\def\eg{{\it e.g.},~}

\newcommand{\dd}{\text{d}}

\newcommand{\beq}{\begin{equation}}
\newcommand{\eeq}{\end{equation}}
\newcommand{\beqn}{\begin{eqnarray}}
\newcommand{\eeqn}{\end{eqnarray}}
\newcommand{\nn}{\nonumber}
\newcommand{\expt}[1]{\left\langle #1 \right\rangle}

\usepackage[nolist]{acronym}
\usepackage{braket}
\usepackage{mleftright}\mleftright
\usepackage{booktabs}
\usepackage{accents}
\usepackage{xfrac}
\usepackage{xspace}
\usepackage{nicefrac}
\usepackage{enumitem}
\usepackage[normalem]{ulem}
\usepackage{soul}
\usepackage{slashed}
\usepackage{microtype}

\newcommand{\ii}{\mathop{}\!\mathrm{i}\!\mathop{}}

\makeatletter
\g@addto@macro\bfseries{\boldmath}
\newcommand*{\defeq}{\mathchoice{\mathrel{\rlap{%
\raisebox{0.24ex}{$\m@th\cdot$}}%
\raisebox{-0.24ex}{$\m@th\cdot$}}%
=}{\mathrel{\rlap{%
\raisebox{0.24ex}{$\m@th\cdot$}}%
\raisebox{-0.24ex}{$\m@th\cdot$}}%
=}{\mathrel{\rlap{%
\raisebox{0.08ex}{\small$\m@th\cdot$}}%
\raisebox{-0.28ex}{\small$\m@th\cdot$}}%
=}{\mathrel{\rlap{%
\raisebox{0.08ex}{\tiny$\m@th\cdot$}}%
\raisebox{-0.28ex}{\tiny$\m@th\cdot$}}%
=}}
\newcommand*{\eqdef}{\mathchoice{=\mathrel{\rlap{%
\raisebox{0.24ex}{$\m@th\cdot$}}%
\raisebox{-0.24ex}{$\m@th\cdot$}}}{%
=\mathrel{\rlap{%
\raisebox{0.24ex}{$\m@th\cdot$}}%
\raisebox{-0.24ex}{$\m@th\cdot$}}}{%
=\mathrel{\rlap{%
\raisebox{0.08ex}{\small$\m@th\cdot$}}%
\raisebox{-0.28ex}{\small$\m@th\cdot$}}}{%
=\mathrel{\rlap{%
\raisebox{0.08ex}{\tiny$\m@th\cdot$}}%
\raisebox{-0.28ex}{\tiny$\m@th\cdot$}}}%
}
\newcommand*{\transpose}{
    {\mathpalette\@transpose{}}%
    }
\newcommand*{\@transpose}[2]{%
    \raisebox{\depth}{$\m@th#1\intercal$}%
}

\makeatother
\title{Cosmological Stasis from Field-Dependent Decay}

\author[a]{Fei Huang}
\author[b]{\!\!, V. Knapp--P\'erez}

\affiliation{\vspace{15pt}$^a$ Department of Particle Physics and Astrophysics, Weizmann Institute of Science, Rehovot 7610001, Israel}
\affiliation{$^b$ Department of Physics and Astronomy, University of California, Irvine, CA 92697--4575 USA}

\emailAdd{fei.huang@weizmann.ac.il}
\emailAdd{vknapppe@uci.edu}

\abstract{Cosmological stasis is a new type of epoch in the cosmological timeline during which the cosmological abundances of different energy components --- such as vacuum energy, matter, and radiation --- remain constant despite the expansion of the universe. 
Previous studies have shown that stasis naturally arises in various scenarios beyond the Standard Model, either through sequential decays of states in large towers or via the annihilation of a single particle species in thermal equilibrium with itself. 
In this work, we demonstrate that stasis can also emerge from the decay of a single particle species whose decay width is dynamically regulated by a scalar field rolling down a Hubble-mass potential.
By analyzing the fixed points of the dynamical system, we identify regions of the parameter space where stasis occurs as a global attractor of cosmic evolution. We also find that, depending on the specific abundance configuration, stasis solutions can manifest as either a stable node with asymptotic behavior or a stable spiral exhibiting intrinsic oscillations.
Furthermore, we present an explicit model for this realization of stasis and explore its phenomenological constraints and implications.}

\begin{document}
\maketitle

\section{Introduction}\label{sec:intro}

It is generally expected that physics beyond the Standard Model (BSM) could manifest in the early stages of the cosmological timeline. 
In turn, the universe's expansion history may also be influenced by BSM effects.
Recently, it has been demonstrated that many BSM theories can give rise to a new kind of epoch dubbed ``stasis'' \cite{Dienes:2021woi,Dienes:2023ziv} in which cosmological abundances of different energy components remain constant over time when the universe continues to expand.
Stasis has been shown to arise in cosmologies involving many possible combinations of such energy components --- components including, for example,
matter, radiation, or vacuum energy \cite{Dienes:2021woi,Dienes:2023ziv,Barrow:1991dn,Dienes:2022zgd,Dienes:2024wnu,Halverson:2024oir,Barber:2024vui,Barber:2024izt}.
For instance, sequential decays of a tower of massive particles can lead to a stasis between matter and radiation \cite{Dienes:2021woi,Dienes:2023ziv,Halverson:2024oir}.
Such towers of states are predicted by many BSM theories, including those involving extra spacetime dimensions, string theories and strongly-coupled gauge theories.
It has also been shown that initial conditions appropriate for stasis can be set up using gravitational interactions \cite{Long:2025wjw}.
Similarly, continuous evaporation of a population of primordial black holes (PBHs) with an extended mass function can also lead to a stasis between matter and radiation wherein the matter component is comprised of the PBHs \cite{Barrow:1991dn,Dienes:2022zgd}.
The matter or radiation component can also be in stasis with vacuum energy if a tower of overdamped scalar fields can transfer its energy consecutively into radiation \cite{Dienes:2023ziv}, or go through a series of overdamped/underdamped transitions \cite{Dienes:2023ziv,Dienes:2024wnu}.
It is even possible that a {\it triple stasis} simultaneously between vacuum-energy, matter, and radiation can also arise if a tower of dynamical scalars experiences successively a combination of underdamping transitions and decays \cite{Dienes:2023ziv}.
In all these scenarios, it has been proven that stasis epochs can emerge naturally as a global attractor of cosmological evolution.
This makes the stasis phenomenon potentially unavoidable within a wide range of parameter space in these BSM cosmologies.

Phenomenologically, modifications to the standard $\Lambda$CDM expansion history may lead to various observable signatures or constraints (see \eg \cite{Batell:2024dsi} for a recent review).
As a highly non-standard cosmological epoch, stasis has the potential to unlock an array of new possibilities.
For example, it has been shown that incorporating a PBH-induced stasis epoch into the cosmological timeline can alter the predictions for the inflationary observables in any model of inflation
and leave distinct imprints on the spectrum of the stochastic gravitational wave background from inflation \cite{Dienes:2022zgd}.
Additionally, a stasis involving vacuum energy can lead to accelerated cosmic expansion, suggesting that the stasis epoch itself has the potential to be the inflationary era \cite{Dienes:2024wnu}.
Indeed, if proven viable, this so-called {\it stasis inflation} may open up a wealth of new theoretical avenues.

In general, the phenomenological aspects listed above depend on the properties of the stasis epoch.
Such properties, in turn, reflect the structure of the underlying theory.
For example, as found in Refs.~\cite{Dienes:2021woi,Dienes:2022zgd,Dienes:2023ziv,Dienes:2024wnu}, 
the abundance configuration during stasis and the duration of stasis are ultimately determined by the polynomial scaling relations for the masses, decay rates, and cosmological abundances across the tower, as well as the hierarchy of relevant fundamental scales.
Recent research employing machine-learning techniques has also extended this analysis to scenarios in which the mass spectra of the states within the tower exhibit exponential growth \cite{Halverson:2024oir}.
Given these examples, it is tempting to claim that evidence for stasis can provide a smoking-gun signal for BSM theories that predict large towers of states.
However, the presence of a tower is not the only way to realize stasis.
For example, it has recently been found that a matter/radiation stasis can also emerge due to annihilations of a {\it single} particle species in thermal equilibrium with itself \cite{Barber:2024vui}.

Indeed, the critical condition for stasis is to have the energy transfer between different components occur at a rate that is balanced against the expansion rate of the universe (which is typically an inverse power-law in the cosmic time).
In Refs~\cite{Dienes:2021woi,Dienes:2022zgd,Dienes:2023ziv,Dienes:2024wnu,Halverson:2024oir,Barrow:1991dn}, such balance is realized by sequential processes (such as particle decay and underdamping transition) across the entire tower of states, while the dynamics of any individual state in the tower do not respect this cancellation in general.
By contrast, in Ref.~\cite{Barber:2024vui}, it is the interplay between the thermally-averaged annihilation rate and the temperature evolution of the matter fields that gives rise to an energy transfer rate that balances the expansion rate.
Nevertheless, it has been shown in Ref.~\cite{Barber:2024izt} that all these scenarios share a common structure of the energy flow.

Up to this point, in stasis scenarios, the decay rate of matter fields were assumed to be constant \cite{Dienes:2021woi,Dienes:2023ziv,Dienes:2024wnu,Halverson:2024oir,Barber:2024izt}. 
In this paper, we point out that one may relax this condition.
Indeed, in many BSM scenarios, couplings for various early-universe processes can acquire non-trivial time dependence by allowing some form of field dependence.
A simple example is to promote the coupling constant $g$ that governs relevant decay processes to a dynamical quantity $g\sim (\phi/\Lambda)^x$.
Indeed, such couplings arise in various scenarios for BSM physics, including  
the Froggatt-Nielsen model for addressing the flavor puzzle \cite{Froggatt:1978nt}, 
the cosmology of flavons (see, \eg \cite{Baldes:2016gaf,Lillard:2018zts,Chen:2019wnk,Borboruah:2024eal}),
early QCD confinement \cite{Ipek:2018lhm},
baryogenesis \cite{Braconi:2018gxo,Croon:2019ugf,Elahi:2020pxl},
dark matter production \cite{Buchmuller:2003is,Berger:2020maa,Batell:2021ofv,Heurtier:2021rko,Cheek:2022yof,Cheek:2024fyc},
and the generation of gravitational waves \cite{Borboruah:2024eal,Blasi:2024vew}.
With the coupling being dynamical, the energy transfer rate due to decay depends on the evolution of the field value $\phi$, which in turn is governed by the form of the potential $V(\phi)$.
In the context of supergravity, the scalar field $\phi$ can couple to energy densities of various components in the universe and thus develop a potential of the form $V\sim H^2\phi^2$ --- a potential commonly referred to as the Hubble mass potential or the Hubble mass term
\cite{Copeland:1994vg,Dine:1995uk,Lyth:2004nx,Kawasaki:2011zi,Kawasaki:2012qm,Kawasaki:2012rs}.
We find that the field evolution under this type of potential can lead to a decay rate that is power-law in time and thus can potentially give rise to a stasis epoch.

The stasis from this type of mechanism is quite different from previous studies in a few ways.
First, the stasis in our scenario is realized through a time-dependent decay rate regulated by field evolution under a particular form of scalar potential instead of certain organized processes across a tower.
Therefore, a {\it single} decaying matter field is enough to produce a prolonged stasis epoch.
Indeed, in the language used in Ref.~\cite{Barber:2024izt}, the mechanism that achieves stasis in our scenario corresponds to an $n=1$ pump, for which we shall provide an explicit particle-physics model.
Second, compared to the stasis from an annihilating species, the decay process in our scenario occurs out of equilibrium and thus does not require thermalization within the matter sector which produces a certain temperature dependence.
Third, while a stasis solution exists for the entire range of the stasis abundance from 0 to 1 in previous scenarios, its existence in our scenario depends on the abundance configuration associated with the corresponding fixed-point solution of the Boltzmann-equation system.
Finally, the stasis in our scenario can be prolonged and even eternal if the coupling $g$ and the scalar potential follow exactly the form that we have prescribed above.
However, to be consistent with current observational data, an exit from stasis is needed.
In the tower-based stasis scenarios, the stasis ends when the decay or transition processes reach the bottom of the tower,
whereas, for stasis in the thermal domain, the end is caused by the inefficiency of annihilations when the temperature drops below a certain minimum.
By contrast, the graceful exit in our scenario can be triggered when either the field-dependent coupling or the Hubble mass term is no longer dominant.
These two different exit mechanisms will lead to different subsequent evolution epochs after stasis.
Fundamentally, these exit mechanisms and, thus, the duration of the entire stasis epoch also reflect the hierarchy between fundamental physics scales in our scenario.

This paper is organized as follows. 
In Sec.~\ref{sec:stasis_condition}, we briefly review the general condition for stasis. 
In Sec.~\ref{sec:field_dependent_decay}, we introduce the field-dependent decay mechanism and show how stasis can emerge from it.
In Sec.~\ref{Sec:StabilityAnalysis}, we study the dynamical properties of the full Boltzmann-equation system and identify regions of the parameter space where the fixed-point solution is stable and can be a global attractor that leads to stasis.
In Sec.~\ref{sec:RealisticScenario}, we present an explicit model for this type of stasis scenario with a mechanism for a graceful exit.
Within this model, we analyze the duration of stasis and the subsequent evolution.
We also analyze constraints from model consistency and phenomenology, and identify regions in the parameter space where the stasis phenomenon is compatible with these constraints.
In Sec.~\ref{sec:conclusion}, we conclude.

\section{Preliminaries and general condition for stasis}\label{sec:stasis_condition}

Let us start with some preliminaries in cosmology and briefly review the condition for stasis.
We shall assume a flat Friedmann-Robertson-Walker (FRW) universe, which contains matter and radiation only.
For any component $i$ (where $i=M,\gamma$ for matter and radiation respectively) with an energy density $\rho_i$ and pressure $p_i$, its equation-of-state parameter $w_i$ and cosmological abundance $\Omega_i$ are defined as
\beqn
w_i &\equiv& \frac{p_i}{\rho_i}\,,\\
\Omega_i &\equiv&\frac{8\pi G}{3H^2} \rho_i\,,
\eeqn
in which $G$ is Newton's gravitational constant, and $H$ is the Hubble parameter.
The Hubble parameter is, in general, a time-dependent quantity. Its evolution is governed by the Friedmann equation
\beq
H^2=\frac{8\pi G}{3}\left(\rho_M+\rho_\gamma\right)\,.
\eeq
Obviously, the Friedmann equation also implies $\Omega_M+\Omega_\gamma=1$.
For this universe, one can define an effective equation-of-state parameter
\beq
w\equiv \frac{p_M+p_\gamma}{\rho_M+\rho_\gamma}=\Omega_M w_M+\Omega_\gamma w_\gamma\,,\label{eq:w_eff}
\eeq
which lies between 0 and 1/3 since $w_M=0$ and $w_\gamma=1/3$.
If $w$ is a constant, the Hubble parameter takes the form
\beq
H=\frac{\kappa}{3t}\,,\label{eq:H_kappa}
\eeq
where $\kappa = 2/(1+w)$ lies between $3/2$ (for radiation domination) and $2$ (for matter domination).

Let us further assume that the matter component is not in thermal equilibrium with radiation but can still transfer its energy to the radiation component.
In general, such energy transfer can be described by a pump term $P^{(\rho)}_{M\gamma}$ in the Boltzmann equations
\beqn
\frac{\dd\rho_M}{\dd t}&=&-3H\rho_M-P^{(\rho)}_{M\gamma}\,,\nn\\
\frac{\dd\rho_\gamma}{\dd t}&=&-4H\rho_\gamma+P^{(\rho)}_{M\gamma}\,.\label{eq:Boltzmann_rho}
\eeqn
Rewriting the above equations in terms of the cosmological abundances, 
it is easy to find that
\beq
    \frac{\dd\Omega_M}{\dd t} 
     = -P_{M\gamma}  
     + H \left( \Omega_M - \Omega_M^2\right)\,,\label{eq:Boltzmann_abundance}
\eeq
where the pump term $P_{M\gamma}$ for the abundances is given by
\beq
P_{M\gamma}\equiv \frac{8\pi G}{3H^2} P^{(\rho)}_{M\gamma}\,,\label{eq:pump_abundance}
\eeq
and we have omitted the differential equation for $\Omega_\gamma$ since $\dd\Omega_\gamma/\dd t = - \dd\Omega_M/\dd t$.
Obviously, if a non-vanishing pump term satisfies and maintains over time the following condition
\beq
   P_{M\gamma}  = H \Omega_M(1 - \Omega_M)\,,\label{eq:pump_StasisCondition}
\eeq
the matter and radiation abundances will stay non-vanishing and constant.
Such phenomenon is referred to as {\it cosmological stasis}.

Since the abundances $\Omega_M$ and $\Omega_\gamma$ are constant during stasis, the effective equation-of-state parameter defined in Eq.~\eqref{eq:w_eff} is also constant, and so is $\kappa$.
As a convention, we shall denote the constant values of $\{\Omega_M,\Omega_\gamma,w,\kappa\}$ during stasis as $\{\overline{\Omega}_M,\overline{\Omega}_\gamma,\overline{w},\overline{\kappa}\}$.
The constant effective equation-of-state parameter implies that $H=\overline{\kappa}/(3t)$.
As a result, the pump term $P_{M\gamma}$ in Eq.~\eqref{eq:pump_StasisCondition} must follow a $1/t$ scaling relation during stasis.

Thus far, we have been completely general on the condition for matter/radiation stasis without imposing any condition on the functional form of the pump term $P_{M\gamma}$.
In Refs.~\cite{Barrow:1991dn,Dienes:2021woi,Dienes:2022zgd,Dienes:2023ziv,Halverson:2024oir}, the pump term $P_{M\gamma}=\sum_\ell \Omega_\ell\Gamma_\ell$ is a linear combination of the decay widths $\Gamma_\ell$ of each individual state $\ell$ in a tower\footnote{Here we generalize the concept of a tower of states to include black holes with an extended mass spectrum.}.
While the decay width $\Gamma_\ell$ of each individual state is a constant in these scenarios,
the weighted sum of all $\Gamma_\ell$'s, with the weights given by the time-dependent abundances $\Omega_\ell(t)$, does generate a $1/t$ time-dependence during stasis.
The pump terms for stasis involving vacuum energy components decaying or transitioning into radiation or matter in Refs.~\cite{Dienes:2023ziv,Dienes:2024wnu} all acquire the $1/t$ dependence in a similar way.

By contrast, in Ref.~\cite{Barber:2024vui}, the $1/t$ scaling behavior in the pump term $P_{M\gamma}= \Omega_M n\expt{\sigma v}$ comes entirely from the annihilation rate of a single species.
Indeed, the interplay between the temperature dependence in the thermally averaged annihilation cross section $\expt{\sigma v}$, the matter number density $n$, and the proper temperature evolution is critical for this realization of stasis.

Interestingly, all these previously studied pump terms can be recast in the following form
\beq
P_{M\gamma} =\Omega_M\Gamma_M(t)\,,\label{eq:pump_eff}
\eeq
where $\Gamma_M$ can be seen as the time-dependent effective energy transfer rate of the entire matter component, which follows the $1/t$ scaling behavior during stasis.
For the pumps mentioned above, we have $\Gamma_M=\sum_\ell \Omega_\ell(t)\Gamma_\ell/\Omega_M$ for the tower-based mechanism and $\Gamma_M=n\expt{\sigma v}$ for the mechanism based on single-species annihilations.

\section{Field-dependent decay as a new mechanism for stasis}\label{sec:field_dependent_decay}
\subsection{Basic ideas}
\label{subsec:BasicIdeas}

In this paper, we present another mechanism that may give rise to stasis in an out-of-equilibrium particle-decay scenario.
We notice that the decay width of a single particle can also be time-dependent if the coupling of the operator that governs its decay is dynamical.
Recalling Eq.~\eqref{eq:pump_eff}, if the decay width itself scales like $1/t$, the collective behavior of a tower will no longer be necessary for sustaining a stasis epoch.

To be concrete, let us consider a scenario in which the matter component is simply a non-relativistic field $\chi$, which is out of equilibrium and can decay into radiation.
The corresponding Boltzmann equations in Eq.~\eqref{eq:Boltzmann_rho} can therefore be written as
\beqn
\frac{\dd\rho_\chi}{\dd t}&=&-3H\rho_\chi-\Gamma_\chi\rho_\chi\,,\nn\\
\frac{\dd \rho_\gamma}{\dd t}&=&-4H\rho_\gamma+\Gamma_\chi\rho_\chi\,,\label{eq:Boltzmann_rhox}
\eeqn
in which $\Gamma_\chi$ is the decay width of $\chi$,
and thus the pump term $P^{(\rho)}_{M\gamma}=\Gamma_\chi\rho_\chi$.
Note that, the two equations above are coupled not only through the decay terms but also through the Hubble parameter $H$ via the Friedmann equation. We shall also assume that the Hubble parameter is dominated by $\rho_\chi$ and $\rho_\gamma$.
Converting the energy densities to the cosmological abundances, Eq.~\eqref{eq:Boltzmann_abundance} becomes
\beq
\frac{\dd\Omega_\chi}{\dd t} = -\Gamma_\chi\Omega_\chi  
     + H \left( \Omega_\chi - \Omega_\chi^2\right)\,.\label{eq:Boltzmann_Ox}
\eeq
Note that, from this point onward, we replace the subscript $M$ with $\chi$ to emphasize that the matter component under consideration is specifically the field $\chi$.
In addition, let us assume that the decay channel is governed by an operator of the form
\beq
g(\phi)\chi\mathcal{O}_{\gamma}\,,
\eeq
where $\mathcal{O}_\gamma$ consists of fields in the radiation bath, and $g(\phi)$ is a coupling that depends on the field value of a scalar field $\phi$.
To be more explicit, we further assume that the field-dependent coupling takes the form
\beq
g(\phi)=\left(\frac{\phi}{\Lambda}\right)^x\,,
\eeq
in which $\Lambda$ represents the UV scale of the effective operator, and $x$ is a positive number.

Obviously, given this type of coupling, the decay width of $\chi$ can be written as
\beq
    \Gamma_\chi= C_\Gamma{\phi^{2x}},\label{eq:Gamma_chi_C}
\eeq
where $C_\Gamma$ is a constant in time.
Its specific time dependence is therefore controlled by the time evolution of $\phi$.
Immediately, we notice that if the equation of motion of $\phi$ admits a stable solution $\phi\sim t^{-1/(2x)}$,
the pump term $P_{\chi\gamma}=\Gamma_\chi\Omega_\chi$ in Eq.~\eqref{eq:Boltzmann_Ox} shall scale like $1/t$ and potentially lead to stasis.
Indeed, this type of pump term corresponds to the $n=1$ case in the generalized pumps $P_{ij}^{(\rho)}=Z\rho_i^n$ described in Ref.~\cite{Barber:2024izt} where $(i,j)$ labels different energy components, and $Z$ is a positive prefactor.
As anticipated, the $n=1$ pump requires a $1/t$ scaling behavior in $Z$ to realize stasis,
which is exactly the scaling relation that we expect the decay width $\Gamma_\chi$ to acquire from the field evolution.
Indeed, as we shall see, this scaling behavior of $Z$ with $t$ arises in a highly non-trivial manner as a consequence of the evolution of an additional dynamical variable --- \ie the field value of $\phi$.

Let us, therefore, examine in what circumstances the scalar field $\phi$ can possess such time dependence.
In what follows, we shall assume that the field $\phi$ always has a negligible cosmological abundance such that it does not significantly affect the background cosmology.
In other words, we shall assume that $\Omega_\chi+\Omega_\gamma \approx 1$, and $\Omega_\phi\ll \Omega_{\chi},\Omega_\gamma$ in our subsequent derivation.
In general, the energy density and pressure of a spatially homogeneous scalar field are given by
\beqn 
\rho_\phi&=\frac{1}{2}\dot{\phi}^2+V\,,\nn\\
p_\phi&=\frac{1}{2}\dot{\phi}^2-V\,,
\eeqn
where $V$ is the potential of $\phi$, and the overhead dot denotes the derivative with respect to the cosmological time $t$.
The equation of motion of $\phi$ is then given by
\beq
\ddot{\phi} +3H\dot{\phi}+\frac{\partial V}{\partial \phi}=0\,.\label{eq:EoM_phi}
\eeq
Assuming that the universe has already entered stasis such that $H =  \overline{\kappa}/(3t)$, where the ``overline'' denotes the constant stasis value of $\kappa$,
we can then insert a special solution $\phi_s(t) = C t^{-1/(2x)}$ into the equation of motion.
It is easy to find that the field dynamics is only compatible with the equation of motion if $\partial V/\partial \phi\sim t^{-1/(2x)-2}$.
It is, therefore, of great interest to speculate what functional form of the potential $V$ can give rise to such field evolution of $\phi$.

One possibility is that the potential $V$ is simply a polynomial function of $\phi$, and thus the power $-1/(2x)-2$ implies that $V\sim \phi^{2+4x}$.
However, since for any non-vanishing positive integer value of $x$ 
the power of $\phi$ in the potential $2+4x\geq 6$,
the insertion of $\phi_s=Ct^{-1/(2x)}$ into the equation of motion will necessarily introduce a relation between the coefficient $C$ and the parameter $\overline{\kappa}$.
Since $\overline{\kappa}$ is a function of the stasis abundances $\overline{\Omega}_M$ and $\overline{\Omega}_\gamma$,
this means that even if a matter/radiation stasis can be achieved via this type of potential, the stasis values, $\overline{\Omega}_M$ and $\overline{\Omega}_\gamma$, are sensitive to the initial value of the special solution $\phi_s(t^{(0)})$ where $t^{(0)}$ denotes the initial time.
For simplicity, we choose not to consider this type of potential in this paper.\footnote{We note that the scenario discussed in Ref.~\cite{Dienes:2024wnu} also exhibits a similar sensitivity wherein the initial field values of the scalar fields in a tower are connected directly to the total energy density of the universe before stasis ends.
The initial field values therefore controls the damping effects via the Hubble friction term.
By contrast, in our context, since $\phi$ is assumed to be a species with negligible abundance before the end of stasis, its initial field value, and thus its initial energy density, has negligible impact to the Hubble parameter.}

On the other hand, if the potential $V$ is quadratic in $\phi$ but at the same time contains a coefficient that is proportional to $1/t^2$ during stasis, 
the insertion of $\phi_s$ into Eq.~\eqref{eq:EoM_phi} will not introduce any relation between the initial field value of $\phi_s$ and the stasis abundance as the coefficient $C$ can be canceled.
It is then reasonable to expect that the potential of the form $V\sim \phi^2/t^2$ can produce a stasis epoch that is insensitive to initial conditions.
We shall, therefore, focus on this well-motivated possibility.

Indeed, assuming $V(\phi) = \lambda_t \phi^2/t^2$ and $H=\overline{\kappa}/(3t)$, in Eq.~\eqref{eq:EoM_phi}, one can obtain a general solution
\beq
\phi(t)=C_1 t^{(1-\overline{\kappa}-\xi)/2} + C_2 t^{(1-\overline{\kappa}+\xi)/2}\,,\label{eq:general_phi}
\eeq
where $\xi=\sqrt{(\overline{\kappa}-1)^2-8\lambda_t}$.
Obviously, for $\xi> 0$, the second term always grows larger, and becomes dominant relative to the first term for sufficiently large $t$.
Identifying the growing solution with $\phi_s\sim t^{-1/(2x)}$, it is straightforward to find that $\xi=\overline{\kappa}-1-1/x$, from which we obtain that
\beq
\overline{\kappa}=1+\frac{1}{2x}+4\lambda_t x\,.\label{eq:kappa_x}
\eeq
We can gain more mileage by noticing that 
\beq
\overline{\kappa}=\frac{2}{1+\overline{w}}= \frac{6}{4-\overline{\Omega}_\chi} \label{eq:kappa_Omegachi}
\eeq
for a universe with $\Omega_\chi+\Omega_\gamma = 1$.
First, inserting Eq.~\eqref{eq:kappa_Omegachi} to 
the condition that $\xi=\overline{\kappa}-1-1/x>0$, we find that
\beq
x>\frac{1}{\overline{\kappa} -1} = \frac{4-\overline{\Omega}_\chi}{2 + \overline{\Omega}_\chi } \,.
\label{eq:x_Omegachi}
\eeq
Since $3/2<\overline{\kappa}<2$ for any $\overline{\Omega}_\chi\in (0,1)$, the above inequality implies that, in order to have $\xi>0$, one needs $x> 1$ at least.
It also implies that only $x\geq 2$ can be consistent with the full range $0<\overline{\Omega}_\chi< 1$.
Thus, $x=2$ can be seen as the {\it minimum} case in which $x$ takes the smallest possible integer value.
Using Eq.~\eqref{eq:kappa_x}, Eq.~\eqref{eq:x_Omegachi} also give rise to a constraint on $x$ and $\lambda_t$ given by
\beq
x \geq \frac{1}{2\sqrt{2 \lambda_t}}\,.
\label{eq:x_relation_lambdat}
\eeq
Physically, this is simply a condition for the existence of a growing solution (with $\xi>0$ in Eq.~\eqref{eq:general_phi}) that scales like $\phi_s\sim t^{-1/(2x)}$.

Second, we can also insert Eq.~\eqref{eq:kappa_Omegachi} to Eq.~\eqref{eq:kappa_x} and obtain a prediction on the matter abundance during stasis,
\beq
\overline{\Omega}_\chi =\frac{4(1-x+8x^2\lambda_t)}{1+2x+8x^2\lambda_t}\,.\label{eq:Omega_chistasis}
\eeq
The physical condition $0<\overline{\Omega}_\chi< 1$ (or equivalently $3/2<\overline{\kappa}<2$ in Eq.~\eqref{eq:kappa_x}) will then put another constraint on the combination of $\lambda_t$ and $x$.

\begin{figure}
    \centering
    \includegraphics[width=0.6\linewidth]{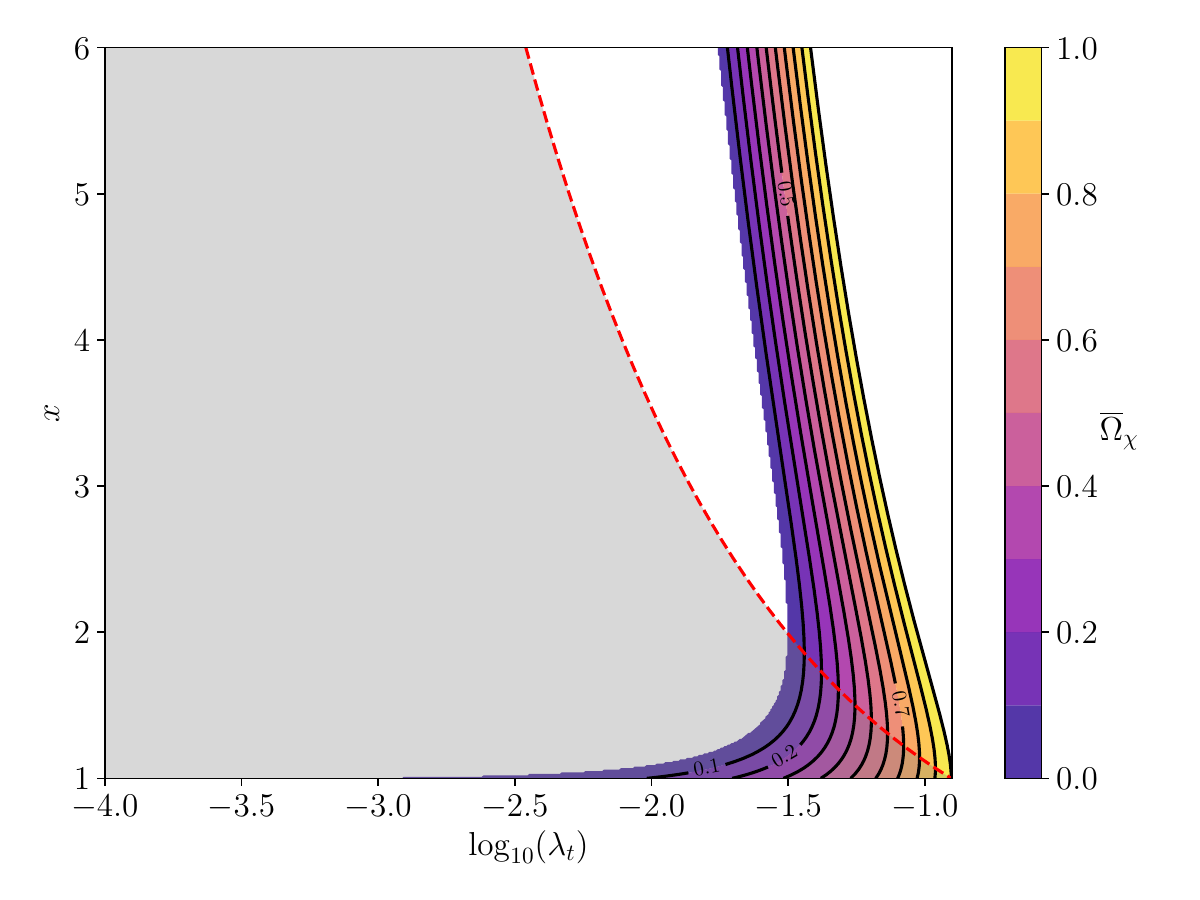}
    \caption{Possible values of the matter abundance $\overline{\Omega}_\chi$ during stasis given by Eq.~\eqref{eq:Omega_chistasis} for different $x$ and $\lambda_t$ values. The gray area bordered by the red dashed line represents the area excluded by Eq.~\eqref{eq:x_relation_lambdat}.}
    \label{fig:ContourOmegaChi}
\end{figure}

In Fig.~\ref{fig:ContourOmegaChi}, we plot contours of $\overline{\Omega}_\chi$ from  Eq.~\eqref{eq:Omega_chistasis} for different values of $x$ and $\lambda_t$. 
The white area represents unphysical values of the $\overline{\Omega}_\chi$, \ie $\overline{\Omega}_\chi<0$ or $\overline{\Omega}_\chi>1$.
The gray area bordered by the red line represents the area excluded by Eq.~\eqref{eq:x_relation_lambdat}. 
As expected, the region in which $1<x<2$ cannot accommodate the full range of $\overline{\Omega}_\chi$.
However, as long as $x\geq 2$, any physical value of $\overline{\Omega}_\chi$ is unconstrained by Eq.~\eqref{eq:x_relation_lambdat}.

\subsection{Motivation for the scalar potential \texorpdfstring{$V\sim \phi^2/t^2$}{V~phi2/t2}}

At first glance, a scalar potential that depends explicitly on the cosmological time $t$ may seem bizarre.
However, we emphasize that the only assumption we made in the previous subsection is that the $1/t^2$ scaling in the potential is exact during the stasis epoch.
The potential can, in general, have a more complicated time-dependence if the background cosmology is not in stasis.
This then allows the possibility of obtaining the appropriate time dependence through other physical quantities instead of via an explicit $t$-dependence.

Fortunately, many physical quantities follow a power-law scaling behavior when the effective equation-of-state parameter $w$ of the entire universe is a constant.
For example, the scale factor $a\sim t^{2/(3+3w)}$, whereas the temperature $T$ of a thermal bath scales like $1/a\sim t^{-2/(3+3w)}$ if entropy is conserved and if there is no change in the effective number of relativistic degrees of freedom.
Likewise, the energy density $\rho_i$ for a species with the equation-of-state parameter $w_i$ scales like $a^{-3(1+w_i)}\sim t^{-2(1+w_i)/(1+w)}$.
We notice that the $1/t^2$ time-dependence emerges if $w_i=w$.
In this case, the energy density $\rho_i$ is proportional to the total energy density and thus to $H^2$.
Indeed, recalling Eq.~\eqref{eq:H_kappa}, $H^2$ possesses the exact $1/t^2$ time-dependence during stasis.
This, therefore, suggests that one can obtain the type of potential, $V\sim \phi^2/t^2$, if the scalar field $\phi$ can couple to energy densities in the universe that are in stasis since, by definition, these energy densities scale the same way as the total energy density in order to maintain constant abundances during stasis.
Fortunately, it turns out that this type of potential is common in the context of supergravity and is referred to as the Hubble mass potential \cite{Copeland:1994vg,Dine:1995uk,Lyth:2004nx,Kawasaki:2011zi,Kawasaki:2012qm,Kawasaki:2012rs}
\beq
V(\phi) =\lambda H^2 \phi^2\,, \label{eq:Hubble_mass}
\eeq
where $\lambda$ encodes the coupling of $\phi$ to different energy densities. 
In general, $\lambda$ can be either positive or negative. 
However, in our context, since we require that $\phi\sim t^{-1/(2x)}$ in order to realize stasis,
this type of behavior cannot be realized if $\phi$ is governed by a Hubble mass potential with a negative $\lambda$.
More importantly, since we shall assume that the Hubble mass term is the dominant contribution to the full scalar potential, a negative $\lambda$ makes the potential unbounded from below --- it would quickly drive the field to infinity which leads to an anti-de Sitter universe.
Therefore, we shall require $\lambda$ to be positive throughout this paper.
Additionally, $\lambda$ should remain constant throughout stasis.
Its relation to $\lambda_t$ is then given by $\lambda=9\lambda _t/\overline{\kappa}^2$.

We note that previous studies have focused on scenarios in which the scalar field obtains the Hubble-mass potential by coupling to the energy density of a single energy component that dominates the universe.
For example, Ref.~\cite{Lyth:2004nx} discussed the Hubble-mass potential during a matter-dominated (MD) era, while Refs.~\cite{Kawasaki:2011zi,Kawasaki:2012qm,Kawasaki:2012rs} studied the same potential in a radiation-dominated (RD) epoch. 
On the other hand, one defining characteristic of cosmological stasis is that it represents a stable mixed-component epoch --- an epoch in which the universe is not completely dominated by any individual energy component.
Therefore, we expect that the $\lambda H^2$ part in the scalar potential includes contributions from the couplings of $\phi$ to energy densities of different components, \ie both $\rho_\chi\sim \Omega_\chi H^2$ and $\rho_\gamma\sim \Omega_\gamma H^2$.
Since these couplings are not necessarily the same,
$\lambda$ can, in general, be a function of the abundances $\Omega_\chi$ and $\Omega_\gamma$.
We shall, therefore, parametrize the coefficient $\lambda$ via the form
\beq
\label{eq:Hubble_mass_f}
\lambda=\lambda_\chi\Omega_\chi+\lambda_\gamma\Omega_\gamma=\lambda'\left[\Omega_\chi+f(1-\Omega_\chi) \right]\,,
\eeq
in which $\lambda'=\lambda_\chi$ and $f\equiv\lambda_\gamma/\lambda_\chi$ parametrizes the relative difference in the coupling strength between the coupling of $\phi$ to $\rho_\chi$ and that to $\rho_\gamma$.
We emphasize again that we have used the condition that $\Omega_\phi$ is negligible compared to the other abundances.
We shall provide an explicit model for this parametrization in Sec.~\ref{sec:RealisticScenario}.

\begin{figure}[t]
\includegraphics[width=0.48\textwidth]{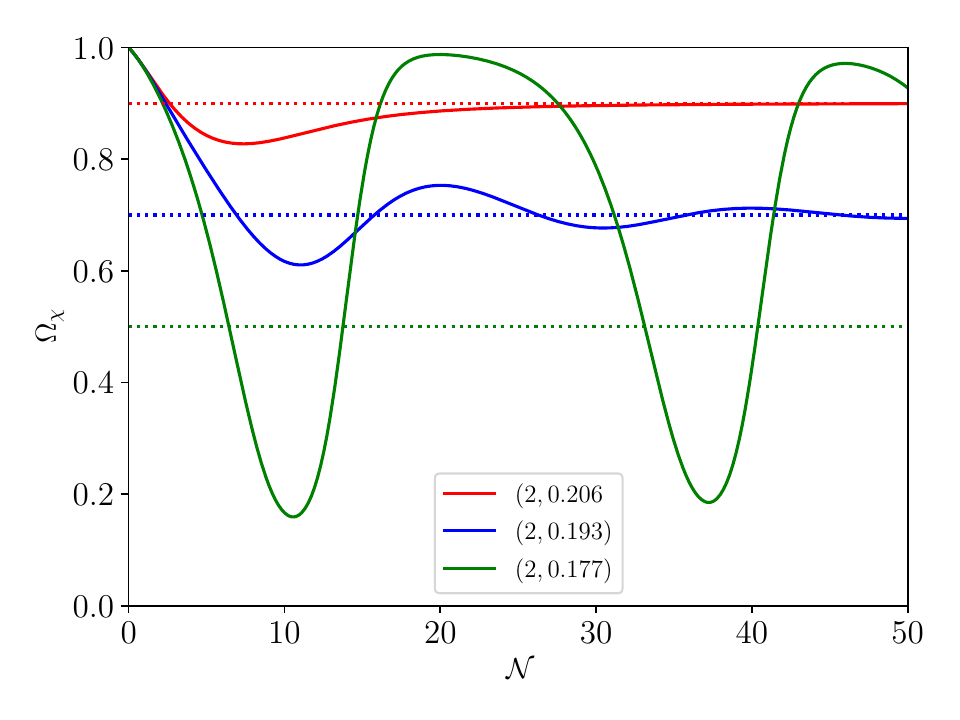}
\includegraphics[width=0.48\textwidth]{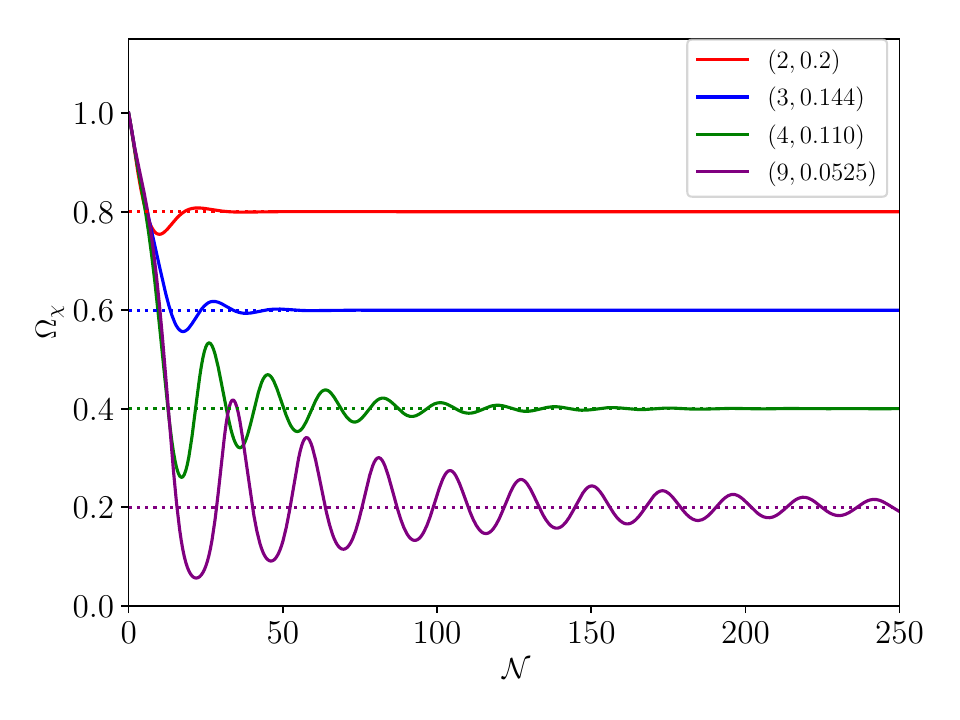}
\caption{The evolution of the matter abundance $\Omega_\chi$ as a function of the number of e-folds $\mathcal{N}$ is shown by the solid curves for $f=1$ and for different values of $(x, \lambda)$ as shown in the legends. 
The value of $\overline{\Omega}_\chi$ in each case is indicated by the dotted horizontal lines.
In the left panel, we fix $x=2$ and show three different types of behavior. 
The cases represented by the red and blue curves correspond to stasis solutions, for which $\Omega_\chi$ either approaches asymptotically to $\overline{\Omega}_\chi$ (red curve)
or oscillates around it with a decreasing amplitude (blue curve).
By contrast, the case represented by the green curve never settles on $\overline{\Omega}_\chi$, and thus stasis is not reached. Instead, $\Omega_\chi$ oscillates with a constant period in $\mathcal{N}$. 
In the right panel, we show how stasis solutions with smaller values of $\overline{\Omega}_\chi$ can be obtained by varying the value of $x$.}
\label{fig:Fig2}
\end{figure}

To proceed further, we collect the full set of differential equations (Eqs.~\eqref{eq:Boltzmann_rhox} and \eqref{eq:EoM_phi}) for our system,
together with the Friedmann equation 
\beq
H^2=\frac{8\pi G}{3}\left(\rho_\chi+\rho_\gamma+\rho_\phi\right)\,,
\label{eq:Friedmann_chigammaphi}
\eeq
which includes the energy density of $\phi$,
and perform a numerical study.
For simplicity, we shall take $f=1$ such that $\lambda =\lambda'$ and does not depend on $\Omega_\chi$ and $\Omega_\gamma$.
In this case, the value of $\overline{\Omega}_\chi$ for specific values of  $\lambda$ and $x$ can be obtained by solving Eq.~\eqref{eq:Omega_chistasis} with Eq.~\eqref{eq:kappa_Omegachi} and $\lambda_t=\overline{\kappa}^2\lambda/9$.
We defer a more general study for the full range of $f$ in the next section.
A few examples are shown in Fig.~\ref{fig:Fig2}, in which we have assumed the following conditions:
\begin{itemize}
    \item the universe is initially matter-dominated with $\Omega_\chi^{(0)}\approx 1$, $\Omega_\gamma=0$, and $\Omega_\phi\ll \Omega_\chi$;
    \item the initial ``velocity'' of the scalar field $\phi$ vanishes, \ie $\left.{d\phi}/{dt}\right|_{t^{(0)}}=0$;
    \item the ratio between the initial values of the decay rate and the Hubble parameter satisfies $\Gamma^{(0)}/H^{(0)} = 0.03$.
\end{itemize}
Notice that we have used the superscript ``$(0)$'' to denote the initial values.
In the left panel of Fig.~\ref{fig:Fig2},
we find that the evolution of $\Omega_\chi$ can in general exhibit three types of behaviors:
\begin{enumerate}
    \item In the case shown by the red curve, $\Omega_\chi$ approaches $\overline{\Omega}_\chi$ asymptotically and eventually converges to the stasis value.
    \item As represented by the blue curve, $\Omega_\chi$ exhibits underdamped oscillatory motion about the stasis value $\overline{\Omega}_\chi$.
    The oscillation amplitude decreases quickly with time $t$, and eventually $\Omega_\chi$ converges to $\overline{\Omega}_\chi$.
    \item For the green curve, $\Omega_\chi$ exhibits undamped oscillation for which the oscillation amplitude becomes essentially constant at later times, and the oscillation period is constant in the variable $\mathcal{N}$. The system never settles to the corresponding $\overline{\Omega}_\chi$.
\end{enumerate}
Indeed, the red and blue curves illustrate scenarios where the solution $\Omega_\chi=\overline{\Omega}_\chi$ is stable and persists over time, though one converges asymptotically, while the other approaches with oscillations.
By contrast, the green curve shows a case in which the $\Omega_\chi=\overline{\Omega}_\chi$ solution is an unstable solution.
Interestingly, these three types of behavior correspond to distinct ranges of $\overline{\Omega}_\chi$,
while the exact ranges depend on the value of $x$.
For our chosen minimal integer value, $x=2$,
these ranges are approximately:
$0.88\lesssim\overline{\Omega}_\chi<1$, where the system exhibits asymptotic convergence towards stasis; $0.6\lesssim\overline{\Omega}_\chi\lesssim 0.88$, where the solution undergoes oscillatory convergence to a stasis value, 
and $0<\overline{\Omega}_\chi\lesssim 0.6$, where undamped oscillations indicate that stasis is never achieved.
By increasing $x$, the region exhibiting the stasis phenomenon can be extended toward smaller values of $\overline{\Omega}_\chi$.
This dependence is shown in the right panel of Fig.~\ref{fig:Fig2}, where stasis with $\overline{\Omega}_\chi\lesssim 0.6$ can emerge by increasing $x$ from 2 to larger integer values.

It is worth pointing out that the dependence of the dynamical behavior on $\overline{\Omega}_\chi$ is very different from all stasis scenarios previously studied \cite{Dienes:2021woi,Dienes:2022zgd,Dienes:2023ziv,Dienes:2024wnu,Halverson:2024oir,Barber:2024vui}.
In these previous works, the stasis solution is always exponentially stable (in the continuous limit) for the full range of physical abundance values such that the approach to the stasis solution is always asymptotic despite potential transient behavior at the beginning.
In the situation where the continuous-limit approximation is no longer valid in a tower-based realization of stasis,
oscillatory behavior may emerge and typically become amplified as time goes on.
Such behavior is caused by the fundamental discreteness of the mass spectrum across the tower.
By contrast, the oscillatory behavior that we observe in this paper originates from the intrinsic property of the solution to the scalar field's equation of motion --- it is due to the non-vanishing imaginary parts in the eigenvalues of the Jacobian matrix, as we shall see in the next section.
Therefore, the oscillatory behavior we observed in this paper is fundamentally different from the oscillatory stasis observed in Ref.~\cite{Dienes:2023ziv}.
Compared to the oscillatory stasis in Ref.~\cite{Dienes:2023ziv},
we also see that, if the oscillation is damped, the time evolution always brings the solution closer to stasis instead of amplifying the deviation.
By contrast, if the oscillation is undamped, this non-stasis solution appears to be stable in both its amplitude and the oscillation frequency measured in $\mathcal{N}$.

\section{Dynamical properties of the system}
\label{Sec:StabilityAnalysis}

Thus far, we have seen that the matter/radiation stasis solution exists
in the limit where the abundance $\Omega_\phi$ of the scalar field is negligible.
We have also seen in the last section that the evolution of $\Omega_\chi$ can exhibit three different types of behaviors, including intrinsic oscillations, which do not necessarily end up in stasis even if a special solution $\Omega_\chi=\overline{\Omega}_\chi$ exists.
To understand the behaviors we have observed, it is therefore important to analyze the dynamical properties of our differential-equation system.

To proceed, let us first write down all the differential equations that govern our dynamical system,
\beqn
\dot{\rho}_\chi &=&-3H\rho_\chi -\Gamma_\chi \rho_\chi\,,\nn\\
\dot{\rho}_\gamma &=&-4H{\rho_\gamma }+\Gamma_\chi \rho_\chi\,,\nn\\
\ddot{\phi}&=&-3H\dot{\phi} - \frac{\partial V}{\partial \phi}\,,\label{eq:dynamical_system_0}
\eeqn
in which $\Gamma_\chi= C_\Gamma{\phi^{2x}}$.
In addition, the Friedmann equation
\beq
\label{eq:Friedmann_Equation}
H^2=\frac{8\pi G}{3}\left(\rho_\chi+\rho_\gamma+\rho_\phi\right)\,,
\eeq
provides a constraint between the Hubble parameter and the energy densities.
To highlight the stasis solution more explicitly, we switch from energy densities $\rho_i$ to the corresponding abundances $\Omega_i$ and also switch the time variable from $t$ to the number of e-folds
\beq
\mathcal{N}\equiv \log a\,,
\eeq
where we have normalized the initial value of the scale factor to be 1.
Using that
$\dd/\dd\mathcal{N}=H^{-1}\dd/\dd t$,
it is easy to show that
\beq
\frac{\dd\Omega_i}{\dd\mathcal{N}}=\frac{\Omega_i}{H}\left( 
\frac{\dot{\rho_i}}{\rho_i}- 2\frac{\dot{H}}{H}\right)=\frac{\Omega_i}{H}\left(\frac{\dot{\rho_i}}{\rho_i}- \frac{\dot{\rho}_{\rm tot}}{\rho_{\rm tot}}\right)\,,
\eeq
where the last term in the parentheses can be re-expressed as
\beq
\frac{\dot{\rho}_{\rm tot}}{\rho_{\rm tot}}=-H\left(3\Omega_\chi+4\Omega_\gamma-\Omega_\phi\frac{\dd\log \rho_\phi}{\dd\mathcal{N}}\right)\,.
\eeq
We therefore obtain
\beqn
\frac{\dd\Omega_\chi}{\dd\mathcal{N}}&=&\Omega_\chi\left(3\Omega_\chi+4\Omega_\gamma -\Omega_\phi\frac{\dd\log \rho_\phi}{\dd\mathcal{N}}-3-\tilde{\Gamma}_\chi\right)\,,\label{eq:dOchidN}\\
\frac{\dd\Omega_\gamma}{\dd\mathcal{N}}&=&\Omega_\gamma\left(3\Omega_\chi+4\Omega_\gamma- \Omega_\phi\frac{\dd\log \rho_\phi}{\dd\mathcal{N}}-4+\frac{\Omega_\chi}{\Omega_\gamma} \tilde{\Gamma}_\chi\right) \,,\label{eq:dOgammadN}\\
\frac{\dd\Omega_\phi}{\dd\mathcal{N}}&=&\Omega_\phi\left(3\Omega_\chi+4\Omega_\gamma-\Omega_\phi\frac{\dd\log \rho_\phi}{\dd\mathcal{N}}+\frac{\dd\log \rho_\phi}{\dd\mathcal{N}}\right)\,,\label{eq:dOphidN}
\eeqn
in which we have defined
\beq
\tilde{\Gamma}_\chi\equiv \frac{\Gamma_\chi}{H}\,.
\eeq
It is easy to verify that $\dd(\Omega_\chi+\Omega_\gamma+\Omega_\phi)/\dd\mathcal{N}=0$ using the Friedmann equation. 
Thus, only two of the differential equations for $\Omega_\chi$, $\Omega_\gamma$, and  $\Omega_\phi$ are independent.
Moreover, to account for the second order differential equation of the scalar field $\phi$ in Eq.~\eqref{eq:dynamical_system_0}, 
we first notice that we can trade $\phi$ for $\tilde{\Gamma}_\chi$ since $\tilde{\Gamma}_\chi\sim \phi^{2x}$.
The first time derivative of $\phi$ is then replaced by
\beq
\frac{\dd\tilde{\Gamma}_\chi}{\dd\mathcal{N}}= \frac{\tilde{\Gamma}_\chi}{2}\left(3\Omega_\chi+4\Omega_\gamma - \Omega_\phi\frac{\dd\log \rho_\phi}{\dd\mathcal{N}}\right)+\Upsilon\,,\label{eq:dGtdN}
\eeq
in which we have defined
\beq
\Upsilon \equiv \frac{\partial \tilde{\Gamma}_\chi}{\partial \phi}\frac{\dd\phi}{\dd\mathcal{N}}=2xC_\Gamma\frac{\phi^{2x-1}}{H}\frac{\dd\phi}{\dd\mathcal{N}}\,.
\eeq
Obviously, the time derivative of $\Upsilon$ will account for the second time derivative of the scalar field $\phi$.
A straightforward calculation yields
\beq
\frac{\dd\Upsilon}{\dd\mathcal{N}}
=\Upsilon\left(3\Omega_\chi+4\Omega_\gamma - \Omega_\phi\frac{\dd\log \rho_\phi}{\dd\mathcal{N}} \right)+\frac{2x-1}{2x}\frac{\Upsilon^2}{\tilde{\Gamma}_\chi}-3\Upsilon -8\lambda'\left[\Omega_\chi+f(1-\Omega_\chi)\right] \tilde{\Gamma}_\chi\,.\label{eq:dUdN}
\eeq

We have now rewritten the equations in the dynamical system in Eq.~\eqref{eq:dynamical_system_0}
in terms of Eqs.~\eqref{eq:dOchidN}, \eqref{eq:dOgammadN}, \eqref{eq:dGtdN} and \eqref{eq:dUdN}.
Since we assume that $\Omega_{\chi},\Omega_\gamma\gg\Omega_\phi$ and $\Omega_\phi\to 0$, to a great approximation, we can neglect any term that is proportional to $\Omega_\phi$, and set $\Omega_\gamma= 1-\Omega_\chi$.
In this limit, the system is reduced to the following differential equations, 
\beqn
\frac{\dd\Omega_\chi}{\dd\mathcal{N}}&=& \Omega_\chi(1-\Omega_\chi-\tilde{\Gamma}_\chi)\,,\nn\\
\frac{\dd\tilde{\Gamma}_\chi}{\dd\mathcal{N}}&=& \frac{\tilde{\Gamma}_\chi}{2}(4-\Omega_\chi)+\Upsilon \,,\nn\\
\frac{\dd\Upsilon}{\dd\mathcal{N}}&=&\Upsilon(1-\Omega_\chi)+\frac{2x-1}{2x}\frac{\Upsilon^2}{\tilde{\Gamma}_\chi}-4x\lambda'\left[\Omega_\chi+f(1-\Omega_\chi)\right]\tilde{\Gamma}_\chi\,.\label{eq:dynamical_system_reduced}
\eeqn
The above form of the dynamical system has several advantages.
First, the system is autonomous as the time variable $\mathcal{N}$ is absent on the right-hand side of all the differential equations.
As a result, the evolution of the system is completely determined by the initial location in the phase space.
Second, 
since both $\tilde{\Gamma}_\chi\sim \Gamma_\chi/H$ and 
$\Upsilon\sim \tilde{\Gamma}_\chi H^{-1}\dd\log\phi/\dd t$ are constant during stasis,
the stasis solution is associated with a fixed point of the dynamical system which satisfies $\dd\Omega_\chi/\dd\mathcal{N}=\dd\tilde{\Gamma}_\chi/\dd\mathcal{N}=\dd\Upsilon/\dd\mathcal{N}=0$.
Interestingly, there are two nontrivial ($\Omega_\chi\neq 0$ or $1$) fixed points as explicit functions of $x$, $f$ and $\lambda'$, which are given by
\beqn
\tilde{\Gamma}_\chi^{\pm}&=&1-\Omega_\chi^{\pm}\,,\nn\\
\Upsilon^{\pm}&=&
-\frac{1}{2}(1-\Omega_\chi^\pm)(4-\Omega_\chi^\pm)\,,\nn\\
\Omega_\chi^{\pm}&=&\frac{2}{1+2x} \left[A(x,f,\lambda')\pm B(x,f,\lambda')\right]\,,
\label{eq:ExplicitOmegaChiStasis}
\eeqn
in which 
\beqn
A(x,f,\lambda')&=&8 (f-1) \lambda'  x^2+x+2\,, \nn\\
B(x,f,\lambda')&=&x \sqrt{9 + 8 \lambda' \left[8 (f-1)^2 \lambda' x^2 + 3f - 2x - 4\right]}\,.
\eeqn
Notice that both fixed points can satisfy the physical condition $0<\Omega_\chi<1$.
For certain combinations of $\{x,f,\lambda'\}$, the two fixed points can even be physical simultaneously.

However, not all fixed-point solutions are stable.
The local stability of our system can be determined by evaluating the eigenvalues of the Jacobian matrix at the fixed points,
\beq
\begin{pmatrix}
\frac{\partial}{\partial \Omega_\chi}\frac{\dd\Omega_\chi}{\dd\mathcal{N}} & \frac{\partial}{\partial \tilde{\Gamma}_\chi}\frac{\dd\Omega_\chi}{\dd\mathcal{N}} & \frac{\partial}{\partial \Upsilon}\frac{\dd\Omega_\chi}{\dd\mathcal{N}}\\
\frac{\partial}{\partial \Omega_\chi}\frac{\dd\tilde{\Gamma}_\chi}{\dd\mathcal{N}} & \frac{\partial}{\partial \tilde{\Gamma}_\chi}\frac{\dd\tilde{\Gamma}_\chi}{\dd\mathcal{N}} & \frac{\partial}{\partial \Upsilon}\frac{\dd\tilde{\Gamma}_\chi}{\dd\mathcal{N}}\\
\frac{\partial}{\partial \Omega_\chi}\frac{\dd\Upsilon}{\dd\mathcal{N}} & \frac{\partial}{\partial \tilde{\Gamma}_\chi}\frac{\dd\Upsilon}{\dd\mathcal{N}} & \frac{\partial}{\partial \Upsilon}\frac{\dd\Upsilon}{\dd\mathcal{N}}\\
\end{pmatrix}_{\left(\Omega_\chi=\Omega_\chi^\pm,\tilde{\Gamma}=\tilde{\Gamma}_\chi^\pm,\Upsilon=\Upsilon^\pm\right)}\,.\label{eq:J_matrix}
\eeq
In general, a fixed point is classified as a node if all eigenvalues of the Jacobian matrix are real; otherwise, it is a spiral point. 
Additionally, if the real parts of all eigenvalues are negative, the fixed point is stable; otherwise, it is unstable.
Moreover, if the eigenvalues have both positive and negative real parts, the fixed point is classified as a saddle point.
We find that the fixed point associated with $\Omega_\chi^-$ can be stable with all the eigenvalues having negative real parts.
However, the other fixed point associated with $\Omega_\chi^+$ always possesses at least one eigenvalue with a positive real part.
Therefore, the latter fixed point is always unstable, and the stasis values can only be identified with the former fixed point, \ie
\beq
\left\{\overline{\Omega}_\chi,\overline{\tilde{\Gamma}}_\chi,\overline{\Upsilon}\right\}=\left\{\Omega_\chi^-,\tilde{\Gamma}_\chi^-,\Upsilon^-\right\}\,.
\eeq

\begin{figure}
\centering
\includegraphics[width=0.49\linewidth]{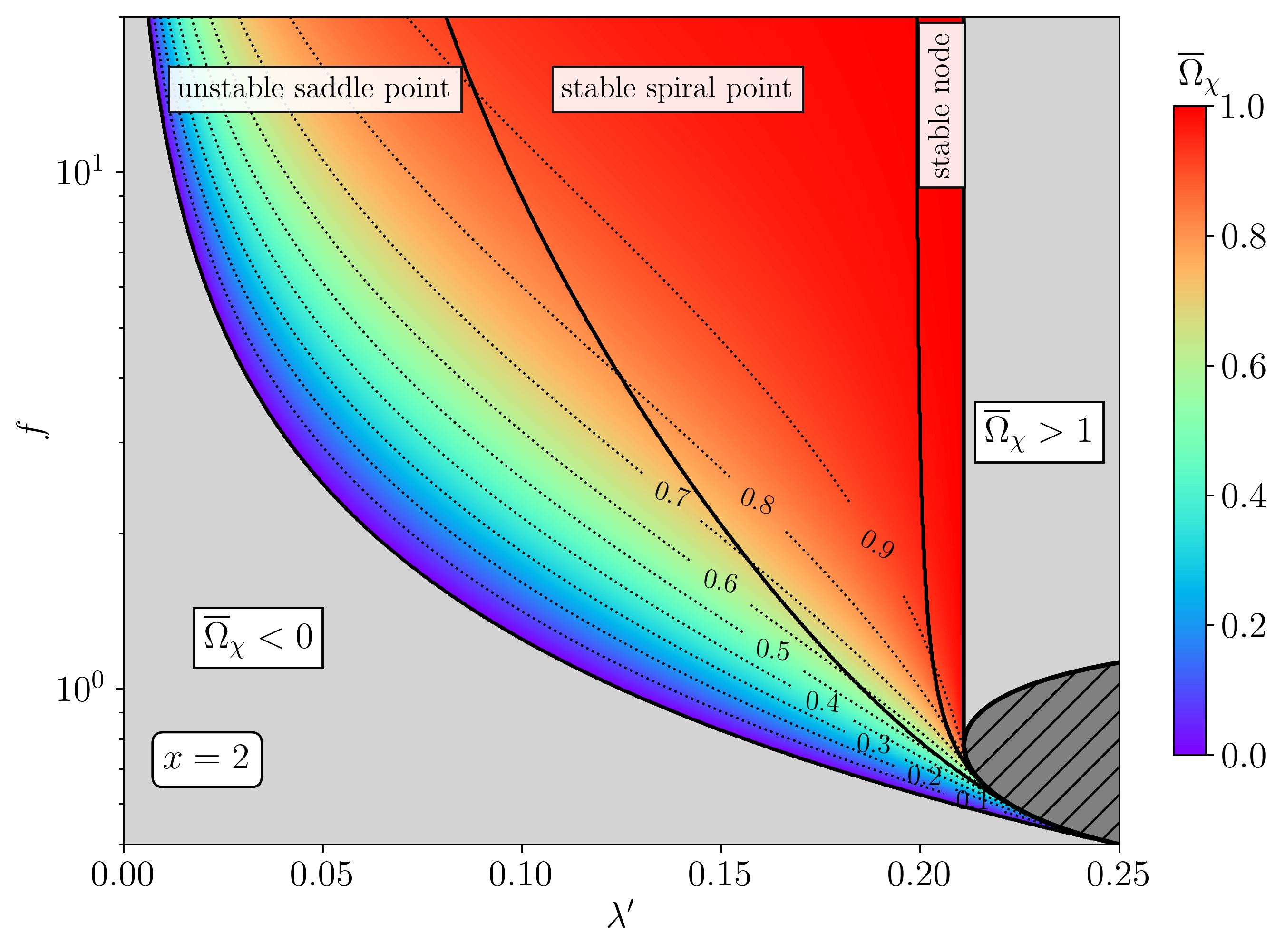}
\includegraphics[width=0.49\linewidth]{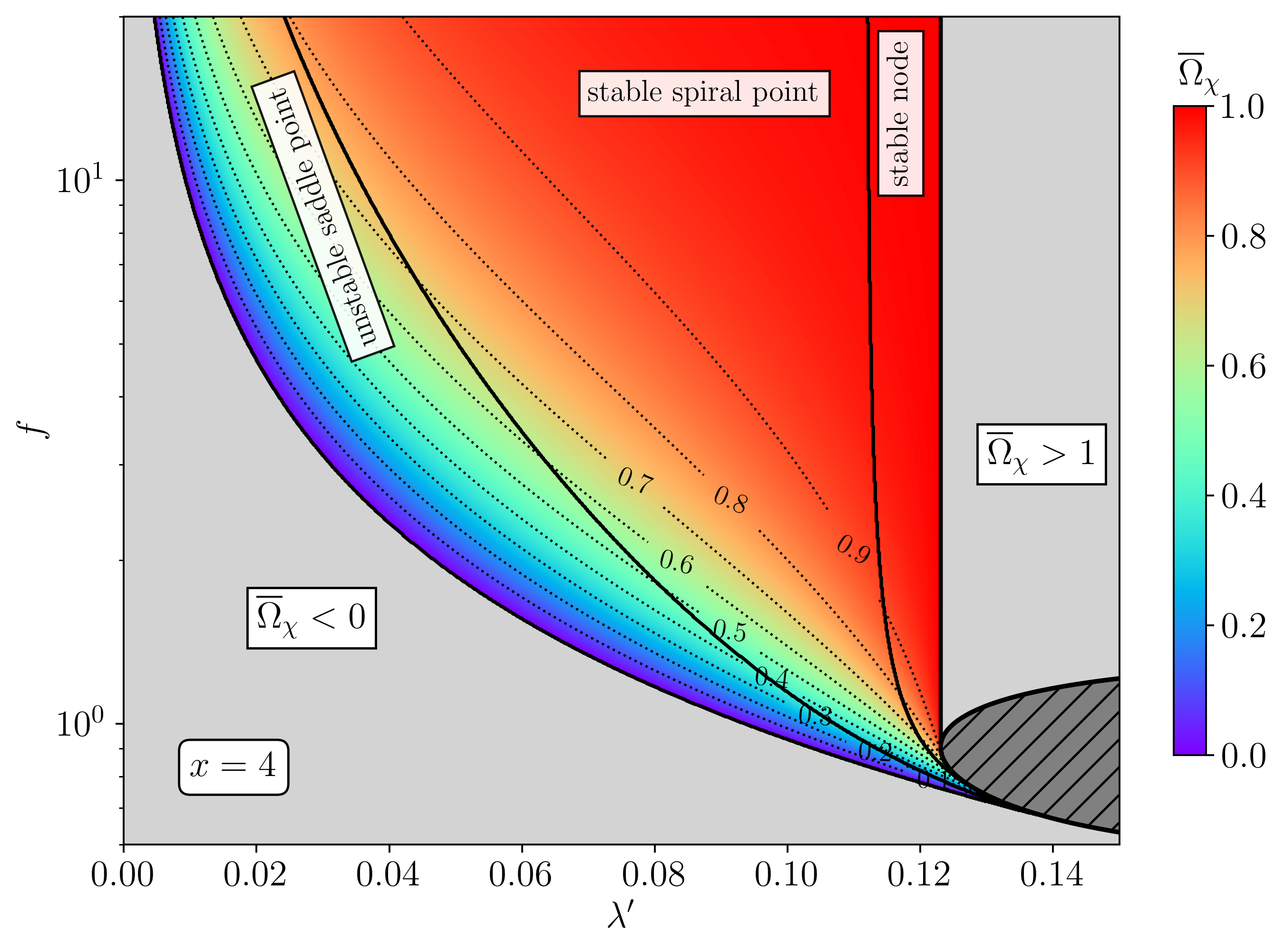}
\caption{In this Figure, we present scans of $\overline{\Omega}_\chi$ and the stability of the fixed-point solution on the $f-\lambda'$ plane while holding $x=2$ and $x=4$ fixed in the left and right panel.
In both panels, $\overline{\Omega}_\chi$ is indicated by the color coding and the dotted black contours.
Notice that the value of $\overline{\Omega}_\chi$ is based on the fixed-point solution $\Omega_\chi^-$.
The gray areas represent regions in which $\overline{\Omega}_\chi$ is unphysical.
In particular, the hatched region corresponds to solutions that are not real. 
The stability of the fixed-point solution is indicated by the text for regions separated by the thick black curves.
}
\label{fig:stability_scan}
\end{figure}

Two scans of $\overline{\Omega}_\chi$ and the stability of the corresponding fixed point across the $\lambda'$-$f$ plane are shown in Fig.~\ref{fig:stability_scan} for $x=2$ and $x=4$.
In both panels, the rainbow-colored region represents the parameter space in which $0<\overline{\Omega}_\chi<1$, as indicated by the thin black dotted curves.
In addition, the thick black curves divide this region into three parts, corresponding to an unstable saddle point, a stable spiral point, and a stable node, respectively.

In both panels, we notice that $f$ is bounded from below, 
while the lower bound (\eg $f>1/2$ for $x=2$) increases slowly as $x$ increases.
For a fixed value of $f$, the range of $\lambda'$ that leads to $0<\overline{\Omega}_\chi<1$ is finite.
However, not all values of $\lambda'$ within this range give rise to a stable fixed point.
The fixed point tends to be unstable for relatively small $\lambda'$.
However, as $\lambda'$ increases, the fixed point first becomes a stable spiral point and then a stable node if $\lambda'$ increases further.
Such behavior is consistent with what we have observed in the left panel Fig.~\ref{fig:Fig2}.
In addition, we also observe that the range of $\lambda'$ that gives rise to a stable fixed point expands as $f$ increases.
However, the range of $\overline{\Omega}_\chi$ that this range covers also shrinks as $f$ increases, as we can see from the dotted black contours --- the stasis solution becomes increasingly matter-like as $f\gtrsim \mathcal{O}(10)$.
Comparing across the two panels of Fig.~\ref{fig:stability_scan}, we also notice that, for a fixed value of $f$, while the range of $\lambda'$ that corresponds to a physical $\overline{\Omega}_\chi$ shrinks as $x$ increases, the range of $\overline{\Omega}_\chi$ that is associated with a stable fixed-point solution expands on the contrary.

The stability of the fixed point guarantees that small perturbations away from the fixed point will not prevent the system from evolving into it.
However, it is also important to investigate whether the universe can evolve into stasis when the initial condition is significantly far away from the fixed point.
We, therefore, analyze the global behavior of the dynamical system by numerically solving the differential equations in Eq.~\eqref{eq:dynamical_system_reduced} with different initial conditions.

\begin{figure}
\centering
\includegraphics[width=0.99\textwidth]{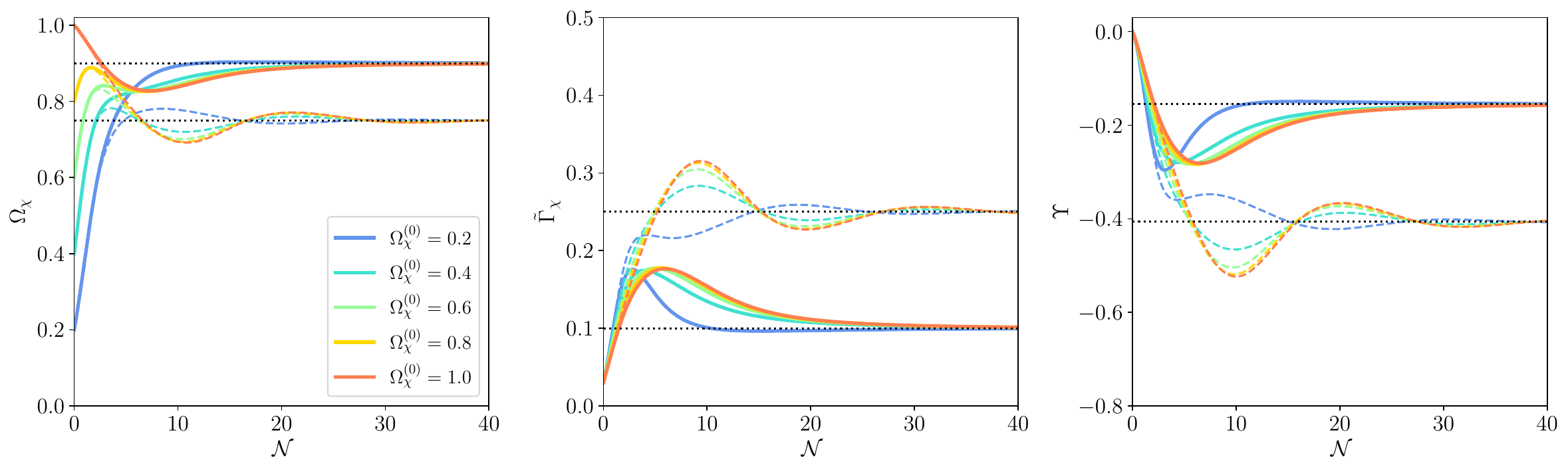}\\
\includegraphics[width=0.99\textwidth]{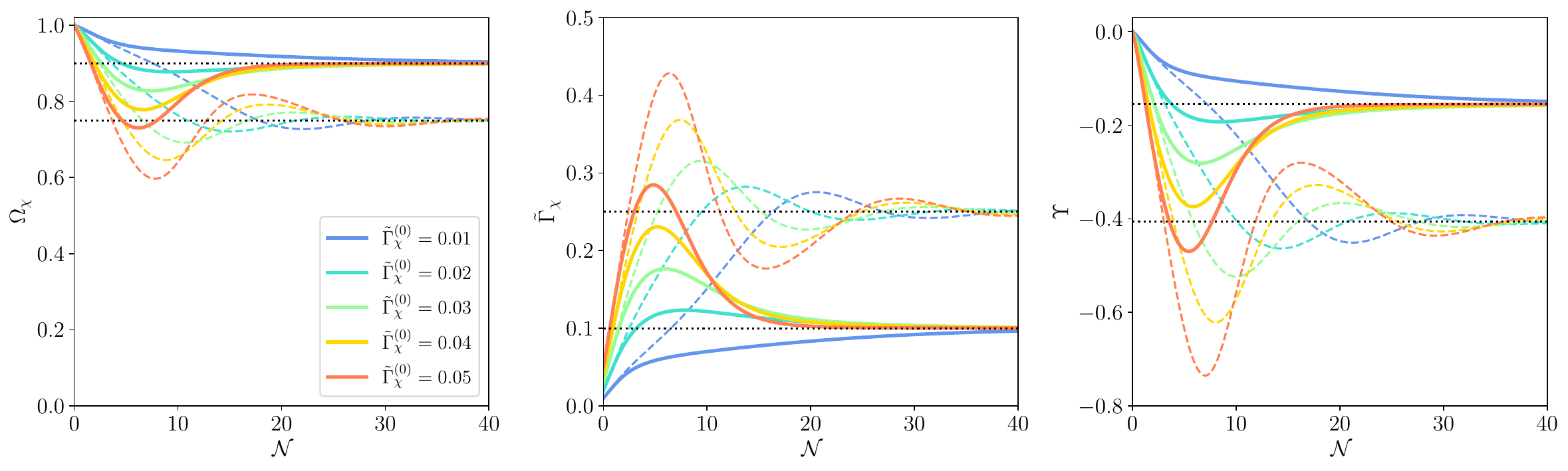}\\
\includegraphics[width=0.99\textwidth]{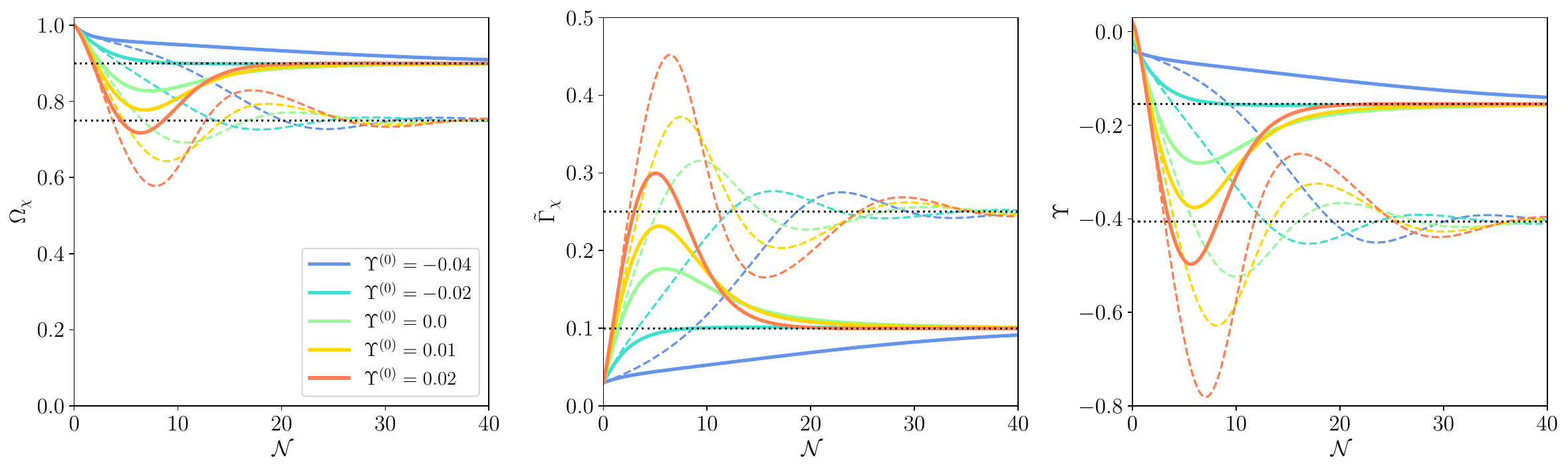}\\
\caption{Approaching stasis from different initial conditions.
In all the plots, we fix $x=2$ and $f=1$ and choose the values of $\lambda'$ such that $\overline{\Omega}_\chi=0.75$ for the dashed curves,
and $\overline{\Omega}_\chi=0.9$ for the solid curves.
These two cases correspond to a stable spiral point and a stable node, respectively.
In each row, we vary one of the initial values $\Omega_\chi^{(0)}$, $\tilde{\Gamma}_\chi^{(0)}$, and $\Upsilon^{(0)}$, while holding fixed the other two.
In particular, we fix $\{\tilde{\Gamma}_\chi^{(0)},\Upsilon^{(0)}\}=\{0.03,0\}$ in the top panels,
$\{\Omega_\chi^{(0)},\Upsilon^{(0)}\}=\{1,0\}$ in the middle panels,
and $\{\Omega_\chi^{(0)},\tilde{\Gamma}_\chi^{(0)}\}=\{1,0.03\}$ in the bottom panels.
}
\label{fig:global}
\end{figure}

To illustrate the global behavior, we fix $\{x,f\}=\{2,1\}$
and present in Fig.~\ref{fig:global} cases with different initial values $\{\Omega_\chi^{(0)}, \tilde{\Gamma}_\chi^{(0)},\Upsilon^{(0)}\}$.
In particular, we choose two benchmark values $\overline{\Omega}_\chi=0.75$
and $\overline{\Omega}_\chi=0.9$ that correspond to a stable spiral point and a stable node, respectively.
We then vary the initial value $\Omega_\chi^{(0)}$ in the top panels, $\tilde{\Gamma}_\chi^{(0)}$ in the middle panels, and $\Upsilon^{(0)}$ in the bottom panels, while holding fixed the other two initial conditions in each row.
For both benchmarks, we can see that the evolution converges to the expected stasis values $\{\overline{\Omega}_\chi, \overline{\tilde{\Gamma}}_\chi,\overline{\Upsilon}\}$.
In particular, all dashed curves in Fig.~\ref{fig:global} exhibit intrinsic oscillatory behavior due to the non-vanishing imaginary parts in the eigenvalues of the Jacobian matrix at the fixed point.
On the other hand, the solid curves all converge to the stasis value asymptotically after some initial evolution that may temporarily drive the evolution away from stasis.
Overall, we find that stasis emerges as a global attractor of the dynamical system as long as the corresponding fixed point is stable,
and different choices of initial conditions only affect how the system approaches stasis.

\FloatBarrier

\section{Toward a realistic scenario: an explicit model}
\label{sec:RealisticScenario}

\subsection{Realizing the Hubble mass term}
\label{subsec:ModelDef}

A Hubble mass term has been considered in the context of supergravity during inflation \cite{Copeland:1994vg,Dine:1995uk}, MD universe \cite{Lyth:2004nx} and RD universe \cite{Kawasaki:2011zi,Kawasaki:2012qm,Kawasaki:2012rs}. We consider a minimal model with only two real scalars $\chi$, $\phi$, and a Dirac fermion $\psi$. 
The scalar $\chi $ is assumed to be an out-of-equilibrium matter field, $\psi$ is in thermal equilibrium with the thermal bath, and $\phi $ is the field that regulates the decay widths of $\chi$.
The Lagrangian is given by
\beqn
    \label{eq:MinimalModelLagrangeDensity}
    \mathcal{L} & =&  \frac{1}{2}\partial_\mu \phi \partial^\mu \phi-\frac{1}{2}m_\phi^2\phi^2 + \frac{1}{2}\partial_\mu \chi \partial^\mu \chi-\frac{1}{2}m_\chi^2\chi^2 + \ii \bar{\psi}\slashed{\partial}\psi - m_\psi \bar{\psi}\psi -y\chi\bar{\psi}\psi \nn \\
    &&  - \lambda_1  \left( \frac{\phi}{M_1}\right)^2 \left( \frac{1}{2}\partial_\mu \chi \partial^\mu \chi+\frac{1}{2}m_\chi^2\chi^2 \right) -\lambda_2\left( \frac{\phi}{M_2}\right)^2 \left( \ii \bar{\psi}\slashed{\partial}\psi + m_\psi \bar{\psi}\psi \right) \nonumber \\
    && -\lambda_3\left( \frac{\phi}{M_3}\right)^2\chi\bar{\psi}\psi \, ,
\eeqn
where $m_\phi$, $m_\chi$, and $m_\psi$ are the classical masses of $\phi$, $\chi$ and $\psi$ respectively, 
$y$ is the Yukawa coupling when $\phi=0$,
$\lambda_i$ denotes the Wilson coefficients of the corresponding non-renormalizable operators, whereas $M_i$ stands for relevant cutoff mass scales. Note that we do not include terms with an odd power of $\phi$ as required to achieve stasis (see Eq.~\eqref{eq:x_Omegachi}). 
We impose this condition by assigning a nontrivial $\mathbb{Z}_2$ charge to $\phi$ while the other fields transform trivially under the $\mathbb{Z}_2$ symmetry.
For simplicity, we shall absorb the Wilson coefficients and write our equations in terms of $\Lambda_i \equiv M_i /\sqrt{\lambda_i}$. 
We remark that, in this notation, $\Lambda_i$ can be larger than the reduced Planck mass $M_P\equiv 1/\sqrt{8 \pi G}$.
However, this does not necessarily mean that the fundamental cutoff scale $M_i$ is larger than the Planck scale.
Instead, it can simply be the result of a small Wilson coefficient $\lambda_i$.

We point out that these terms may come from a K\"ahler potential of the form $\mathcal{K} = \Phi_i \Phi_i^{\dagger}\left(1 +\lambda_i{\varphi \varphi^{\dagger}}/{M_i^2}\right)$, where $i=1,2$,
and $\Phi_{1,2}$ represents the superfield associated with $\chi$ and $\psi$ respectively, and $\varphi$ represents the superfield that contains the scalar $\phi$. 
Hence, terms both in the potential and kinetic part of the Lagrangian get a factor of roughly $\left(1 + \lambda_i {\phi^2}/{M_i^2}\right)^{-1} \approx \left(1 - \lambda_i {\phi^2}/{M_i^2}\right)$ for $\phi \ll M_i$, where $\phi \equiv \left| \varphi \right|$, and we use the notation that $\varphi$ represents both the superfield and its scalar component. We refer readers to \cite{Copeland:1994vg,Dine:1995uk,Lyth:2004nx} for further details.

The Hubble mass potential of the scalar field $\phi$ originates from the couplings of $\phi$ to the energy densities of $\chi$ and $\psi$. 
In particular, the couplings between $\chi$ and $\phi$ in  the second line of Eq.~\eqref{eq:MinimalModelLagrangeDensity} gives rise to
\beq
\label{eq:rhoChi}
    V_1(\phi, \rho_\chi) \equiv \left( \frac{\phi}{\Lambda_1}\right)^2 \left( \frac{1}{2}\partial_\mu \chi \partial^\mu \chi+\frac{1}{2}m_\chi^2\chi^2 \right) =  \left( \frac{\phi}{\Lambda_1}\right)^2 \rho_\chi\,.
\eeq
On the other hand, since $\psi$ is in thermal equilibrium with the thermal bath, operators involving $\psi$ acquire thermal expectation values.
In particular, the kinetic term of $\psi$ gives a contribution to the effective potential of $\phi$ proportional to the radiation energy density \cite{Kawasaki:2012qm}, which is given by
\beqn
\label{eq:kineticThermalTerm}
    V_{2, K}(\phi , \rho_\gamma) \equiv  \left(\frac{\phi}{\Lambda_2}\right)^2\braket{\ii \bar{\psi}\slashed{\partial}\psi}_{\text{th}} \approx 2\frac{m_{\psi, \text{th}}^2 T^2}{12}  = \frac{5 \zeta g^2}{\pi^2 g_{\star}(T)}\left(\frac{\phi}{\Lambda_2}\right)^2 \rho_\gamma\, ,
\eeqn
where $\expt{\dots}_{\rm th}$ denotes the thermal expectation value.
In the second step, the factor of $2$ comes from the fact that $\psi$ is a $4$-component fermion,
and $m_{\psi, \rm{th}}^2 = \zeta g^2 T^2$ is the thermal mass with $\zeta\lesssim \mathcal{O}(1)$ from thermal field theory and $g \ll 1$ being the coupling of $\psi$ to the thermal bath.\footnote{Notice that we have assumed that $m_{\psi, \rm th}\gg m_\psi$ in Eq.~\eqref{eq:kineticThermalTerm}.
The form of the thermal mass implies that $T\gg m_{\psi, \rm th}$. This is consistent with the assumption that $\psi$ is relativistic
and further implies that $T\gg m_\psi$.}
In the last step, we have also used that $\rho_{\gamma} = \frac{\pi^2}{30} g_{\star}(T) T^4$, where $g_{\star}$ counts the effective number of relativistic degrees of freedom. 
In addition, the coupling between $\phi$ and the mass term of $\psi$ in the second line of Eq.~\eqref{eq:MinimalModelLagrangeDensity} generates a thermal contribution
\cite{Dolan:1973qd}
\beqn
\label{eq:Vmpsi}
    V_{2,M}(\phi, \rho_\gamma )  &\equiv&  \frac{\left(m_\psi^2+m_{\psi, \text{th}}^2\right)}{12}\frac{\phi^4}{\Lambda_2^4}T^2  =  \frac{m_\psi^2}{12 \pi} \sqrt{\frac{30 \rho_\gamma}{ g_\star(T)}}\frac{\phi^4}{\Lambda_2^4} + \frac{5 \zeta g^2}{2\pi^2 g_{\star}(T)}\frac{\phi^4}{\Lambda_{2}}\rho_\gamma\nn\\
    &\approx& \frac{5 \zeta g^2}{2\pi^2 g_{\star}(T)}\frac{\phi^4}{\Lambda_2}\rho_\gamma\,.
\eeqn
Notice that in the above equation
we have assumed that $m_\psi \ll T$. 
In total, the effective potential of $\phi$ can be written as
\beqn
    V(\phi) & = & \frac{1}{2}m_\phi^2\phi^2 + V_1(\phi , \rho_\chi ) + V_2(\phi , \rho_\gamma )\nn \\
    & \approx & \frac{1}{2}m_\phi^2\phi^{2} + \left[ \frac{\rho_\chi}{\Lambda_1^2} + \frac{ 5 \zeta g^2\rho_\gamma}{\Lambda_2^2\pi^2 g_{\star}(T)} \right]\phi^2  +  \frac{5\zeta g^2}{2\pi^2 g_\star(T)}\frac{\rho_\gamma}{\Lambda_3^4} \phi^4  \nn \\
    & = & \frac{1}{2}m_\phi^2\phi^{2} +  \lambda^{\prime} H^2\left( \Omega_\chi + f\Omega_\gamma\right) \phi^2 +  \frac{5\zeta g^2}{2\pi^2 g_\star(T)}\frac{\rho_\gamma}{\Lambda_3^4}\phi^4\, ,\label{eq:Veff}
\eeqn
where $V_2=V_{2,K}+V_{2,M}$.
We have also made the following identifications in the third line,
\beq
\lambda^{\prime} \equiv    \frac{3M_P^2}{\Lambda_1^2}\,, ~~~~~f \equiv \frac{ 5 \zeta g^2}{\pi^2 g_\star(T)}\frac{\Lambda_1^2}{\Lambda_2^2}\,.\label{eq:para_identification}
\eeq
Obviously, the second term of in the last line of Eq.~\eqref{eq:Veff} is exactly the Hubble mass potential proposed in Eqs.~\eqref{eq:Hubble_mass} and~\eqref{eq:Hubble_mass_f}. 
With $\phi \ll \Lambda_i$ as required by the effective field theory, we can see that the third term in the last line of Eq.~\eqref{eq:Veff} is always subdominant compared to the Hubble mass term.
As a result, we find that the effective potential of $\phi$ is dominated by the Hubble mass term if $m_\phi\ll H$, \ie
\beq
V(\phi)\approx \lambda^{\prime} H^2\left( \Omega_\chi + f\Omega_\gamma\right) \phi^2\,.
\eeq

With the Lagrangian in Eq.~\eqref{eq:MinimalModelLagrangeDensity}, the non-relativistic matter field $\chi$ mainly decays through three different channels:
the 2-body channel $\chi\to \bar\psi+\psi$,
the 4-body channel $\chi\to \bar\psi+\psi+\phi+\phi$,
and another 2-body channel 
$\chi\to \phi+\phi$ obtained by closing the fermion legs into a loop. Notice that, due to the $\mathbb{Z}_2$ symmetry, $\phi$ can only be produced in pairs.
The corresponding decay widths in the limit $m_\chi\gg m_\psi, m_\phi$ are given by
\beqn
\Gamma_{\chi\to\bar\psi\psi} &=&  \left(y+\frac{\phi^2}{\Lambda_3^2}\right)^2 \frac{m_\chi}{8 \pi}\, ,\label{eq:DecayRateExplicitModel}\\
\Gamma_{\chi\to\bar\psi\psi\phi\phi}&=&\frac{m_\chi^5}{294912\pi^5\Lambda_3^4}\,,\label{eq:4-body}\\
\Gamma_{\chi \to \phi\phi} &=& \frac{1}{8192 \pi^5}\left( \frac{\Lambda_{\rm min}}{\Lambda_3}\right)^2\left( \frac{m_{\psi}  + m_{\psi,\rm th}}{m_\chi}\right)^2  m_\chi\,,\label{eq:2-bodyClosedFermion}
\eeqn
in which $\Lambda_{\rm min}\equiv \min\left\{ \Lambda_1, \Lambda_2, \Lambda_3 \right\}$ is the cut-off scale of the fermion loop.
Note that, at one loop, the Yukawa term and its field-dependent counterpart give rise to a thermal correction $m_{\chi,\rm th}^2\approx  \frac{1}{4}\left[y+(\phi/\Lambda_3)^2\right]^2T^2$ 
to the mass of $\chi$ particles
(see \eg the last line of Eq.~(28) in Ref.~\cite{Carrington:1991hz}).
The consistency of our scenario therefore requires that $m_\chi\gg m_{\chi, \rm th}$.
To be conservative, we shall require that this condition is satisfied even for the highest temperature $T_{\rm max}$ attained in our numerical results. 
We emphasize that, since the temperature $T$ decreases with time monotonically except for the transient epoch before reaching $T_{\rm max}$, a violation of this consistency constraint does not necessarily prevent stasis from occurring.
Instead, it may only abridge the duration of stasis.
Nevertheless, we have checked all our numerical results such that this consistency constraint is always satisfied. 
We therefore do not include $m_{\chi, \rm th}$ in Eqs.~\eqref{eq:DecayRateExplicitModel}, \eqref{eq:4-body} and~\eqref{eq:2-bodyClosedFermion}.
On the other hand, the mass of $\psi$ particles in Eq.~\eqref{eq:2-bodyClosedFermion} does contain its thermal corrections since we have assumed that the thermal mass of $\psi$ is dominant. 
As a sanity check, we find that $m_{\chi}\gg T$ at all times in all our cases. Thus, the decay widths in the equations above are always kinematically accessible.

Apparently, for $y\ll (\phi/\Lambda_3)^2$, the decay width $\Gamma_{\chi\to\bar\psi\psi}$ reduces to the form we propose in Eq.~\eqref{eq:Gamma_chi_C}, in which we can identify the parameter $C_\Gamma =m_\chi/(8\pi\Lambda_3^4)$.
In addition, to achieve stasis, the field-dependent decay mechanism also requires that $\Gamma_{\chi\to\bar\psi\psi\phi\phi}$ and $\Gamma_{\chi \to \phi\phi}$ are subdominant compared to $\Gamma_{\chi\to\bar\psi\psi}$.
Taken together, these requirements, in turn, lead to conditions on the field value $\phi$ that
\beqn
\phi &\gg& \sqrt{y} \Lambda_3\,, \nn\\
\phi &\gg& 0.023 m_\chi\,,\nn\\
\phi &\gg& 0.056\times \left(\frac{m_\psi+m_{\psi,\rm th}}{m_\chi}\right)^{1/2}\sqrt{\Lambda_3\Lambda_{\rm min}}\,.
\eeqn
Note that, due to the thermal correction $m_{\psi,\rm th}\sim gT$, the third line of the above inequality is time-dependent.
However, while $\phi\sim t^{-1/4}$ during stasis, the thermal correction $m_{\psi, \rm th}\sim T \sim \rho_\gamma^{1/4}\sim \overline{\Omega}_\gamma^{1/4} H^{1/2}\sim t^{-1/2}$.
As a result, in the limit $m_{\psi, \rm th}\gg m_\psi$, both sides of the corresponding inequality scale with time in the same manner.
Indeed, in this limit, both $\Gamma_{\chi\to\psi\bar\psi}$ and $\Gamma_{\chi\to\phi\phi}$ scale like $H$ during stasis .
Therefore, when the thermal contributions dominate the mass of $\psi$, to ensure the validity of this inequality,
it is sufficient to require that  
$\Gamma_{\chi\to\bar\psi\psi} \gg \Gamma_{\chi \to \phi\phi}$ at the maximum temperature during stasis.
We have verified that this condition holds in all numerical results presented in this work. 
However, as we shall show in Sec.~\ref{subsec:DurationStasis}, 
the tree-level contribution to $\Gamma_{\chi\to\bar\phi\phi}$ can impact the duration of stasis since $\Gamma_{\chi\to\bar\psi\psi}$ decreases with time and can become comparable to $\Gamma_{\chi\to\phi\phi}$ at later times.

Moreover, the coupling terms in the second and third lines of Eq.~\eqref{eq:MinimalModelLagrangeDensity} can also mediate the 2-to-2 process $\phi+\phi\leftrightarrow\bar\psi+\psi$, and the 2-to-3 process $\bar\psi+\psi\leftrightarrow\phi+\phi+\chi$ and $\phi+\phi\leftrightarrow\bar\psi+\psi+\chi$.
To be consistent with our assumption that both $\phi$ and $\chi$ are out of equilibrium, we need all these interaction rates to be smaller than the expansion rate $H$.
Such constraints can be estimated using simple dimensional analysis.
Notice that, the 2-to-2 process is not only mediated by the operator $(m_\psi/\Lambda_2^2)\phi^2\bar\psi\psi$, but also by $(\expt{\chi}/\Lambda_2^2)\phi^2\bar\psi\psi$, in which $\expt{\chi}\sim \sqrt{\rho_\chi}/m_\chi$ also generates a field-dependent coupling.
These two operators generate contributions of the order  $m_\psi^2/\Lambda_2^4$ and $\rho_\chi/(m_\chi^2\Lambda_3^4)$ to the thermally averaged cross section $\expt{\sigma_{\bar\psi\psi\to\phi\phi}v}$, respectively. 
Using that the number density $n_\psi\sim T^3$, we obtain the following two conditions
\beq
T\lesssim \frac{\Lambda_2^4}{m_\psi^2 M_P \sqrt{\Omega_\gamma}}
~~~\text{and}~~~
  T\lesssim \left(\frac{\sqrt{\Omega_\gamma}}{\Omega_\chi} \frac{m_\chi^2\Lambda_3^4}{M_P}\right)^{1/5}\,.\label{eq:2to2}
\eeq
Notice that in the above estimate, we have used that $H\sim T^2/(\sqrt{\Omega_\gamma} M_P)$.
In addition, for the 2-to-3 processes $\bar\psi+\psi\leftrightarrow\phi+\phi+\chi$ mediated by $(\chi/\Lambda_2^2)\phi^2\bar\psi\psi$,
the thermally averaged cross section $\expt{\sigma_{\bar\psi\psi\to\phi\phi\chi}v}\sim T^2/\Lambda_3^4$, where $T$ is the typical momentum of the $\psi$ particles.
After simple algebra, we obtain that
\beq
T\lesssim \left(\frac{1}{\sqrt{\Omega_\gamma}}\frac{\Lambda_3}{M_P}\right)^{1/3} \Lambda_3\,.\label{eq:2to3}
\eeq
A similar analysis can be also performed for the other 2-to-3 process by assuming $\phi$ is in equilibrium, which yields the same result.
We emphasize that the constraints in Eqs.~\eqref{eq:2to2} and \eqref{eq:2to3} are not difficult to satisfy, and they are met in all our numerical results presented in this paper.

With the above constraints taken into account, we have thus shown that the basic ingredients for realizing stasis through the mechanism we introduced in previous sections can be realized by an explicit model with the Lagrangian in Eq.~\eqref{eq:MinimalModelLagrangeDensity}.  

\subsection{Exiting stasis}\label{sec:exit_stasis}
Once entered, the stasis epoch is, in principle, eternal if the Hubble mass term and the field-dependent coupling continue to be dominant in the scalar potential and in the decay width $\Gamma_\chi$.
However, in a viable cosmological scenario, an exit from stasis is necessary to allow consistency with observational constraints on the expansion history of our universe. 
In particular, the universe needs to reheat and become radiation-dominated with a temperature above $\mathcal{O}({\rm MeV})$ in order to ensure a successful Big Bang Nucleosynthesis (BBN).

In general, for our realization of stasis, a deviation from the scaling relations required by stasis can occur for three different reasons:
1) the decay width $\Gamma_\chi$ is no longer controlled by the field-dependent coupling;
2) the Hubble mass term becomes subdominant in the scalar potential $V(\phi)$;
3) the Hubble mass term becomes invalid.
In our model, these three conditions are triggered by the existence of the Yukawa coupling $y$, the inclusion of the classical mass $m_\phi$, and the violation of the condition $m_{\psi,\rm th}\gg m_\psi$, respectively.
In addition, we note that the temperature dependence of $g_\star(T)$, which typically manifests at the electroweak and QCD phase transition scales in the Standard Model, can also influence stasis through variations in $f$.
However, since $g_\star(T)$ decreases monotonically as the temperature drops, Eq.~\eqref{eq:para_identification} implies that $f$ can only increase with time.
Therefore, as shown in Fig.~\ref{fig:stability_scan}, 
such variations only modify the value of $\overline{\Omega}_\chi$ without destablizing stasis.

While all three exiting mechanisms are possible and well-motivated, due to the theoretical uncertainty in the scalar potential $V(\phi)$ after the Hubble mass term becomes invalid,
we shall, for simplicity, not consider the third exiting mechanism.
Instead, we shall ensure that the Hubble mass term is always valid before the field $\phi$ starts oscillating, and only consider the exit of stasis caused by the Yukawa coupling $y$ and the classical mass $m_\phi$.

In the first exiting mechanism, the Yukawa coupling $y$ eventually becomes comparable to --- and later dominates --- the field-dependent coupling $\phi^2/\Lambda_3^2$ as the scalar field $\phi\sim t^{-1/4}$ keeps decreasing during stasis.
The same argument also applies even if the scalar field $\phi$ has started oscillating since the time average $\expt{\phi^2}_t$ also decreases with time.
Let us define an instant $t_y$ as the moment when the Yukawa coupling is equal to the field-dependent coupling, \ie
\beq
    \frac{\phi(t_y)^2}{\Lambda_3^2} = y \, .
    \label{eq:ty0}
\eeq
After this moment, the decay rate in Eq.~\eqref{eq:DecayRateExplicitModel} stops following the Hubble rate and 
quickly settles to its minimum value
\beq
    \Gamma_{\chi}^{\mathrm{min}} = y^2 \frac{m_\chi}{8 \pi} \,.
    \label{eq:GammaMin}
\eeq
The matter field $\chi$ then decays exponentially into radiation as the Hubble rate falls below the minimum decay width $\Gamma_{\chi}^{\rm min}$,
causing the universe to transition into an RD epoch.
In the case in which the end of stasis is indeed triggered by the Yukawa coupling $y$,
we can obtain an estimate of $t_y$ from Eq.~\eqref{eq:ty0} that
\beq
    t_y = \frac{\phi^4(t_*)}{\Lambda_3^4 y^2}t_*\,,
    \label{eq:ty}
\eeq
where $t_*$ is an arbitrary fiducial time within stasis, and we have used that $\phi \sim t^{-1/4}$ during stasis.

For the second exiting mechanism,
as the Hubble parameter decreases with time,
the classical mass term in Eq.~\eqref{eq:Veff} eventually becomes dominant,
and then, the scalar field starts oscillating.
The onset of oscillation is by convention defined as the instant at which $3 H(t_{\mathrm{osc}})= 2 m_\phi$. 
Parametrizing the Hubble parameter as $H=\kappa/(3t)$, we obtain that
\beq
    t_{\mathrm{osc}} = \frac{\kappa}{2 m_\phi}\, .
    \label{eq:tosc}
\eeq
Notice that $\kappa$ can, in general, be time-dependent and vary between $3/2$ and $2$ as the universe exits stasis.
Nevertheless, if stasis ends at $t_{\rm osc}$, we can approximate $\kappa$ as $\overline{\kappa}$.

Once the oscillation starts, $\phi$ starts to behave like non-relativistic matter with 
$\rho_\phi\sim \expt{\phi^2}_t \sim a^{-3}$.
In the meantime, if the decay width of $\chi$ is still dominated by the field-dependent coupling, it is easy to find that $\Gamma_\chi \sim \expt{\phi^4}\sim a^{-6}$. 
This implies that, 
while the decay rate and the Hubble expansion rate are of similar order during stasis,
oscillations of $\phi$ will rapidly drive the decay rate below the Hubble rate.
As a result, the matter abundance $\Omega_\chi$ will start to increase relative to the radiation abundance $\Omega_\gamma$,
and the universe will enter a matter-dominated epoch after stasis ends at the oscillation time $t_\mathrm{osc}$.

In summary, since the domination of $y$ and the oscillation of $\phi$ will eventually both occur, the order of the two moments $t_y$ and $t_{\rm osc}$ tells us which mechanism triggers the end of stasis.
We, therefore, have two different cases:
\begin{itemize}
    \item Case I: $t_y < t_{\mathrm{osc}}$. The universe enters a RD era after stasis ends at $t_y$.
    \item Case II: $t_{\mathrm{osc}} < t_y$. The universe enters an MD era after stasis ends at $t_{\mathrm{osc}}$.
\end{itemize}

In both cases, the universe eventually enters a RD era around the reheating time $t_{\mathrm{RH}}$ at which 
\beq
\Gamma_{\chi}=H(t_{\mathrm{RH}})\, .
    \label{eq:tRD0}
\eeq 
Approximating the decay rate as $\Gamma_{\chi}^{\mathrm{min}}$ and the Hubble rate as $H(t_{\rm RH})=1/(2t_{\rm RH})$ at the reheating time,
we obtain that
\beq
    t_\mathrm{RH} = \frac{4 \pi}{y^2 m_\chi}\, .
    \label{eq:tRD}
\eeq

\begin{figure}[t]
\centering
\includegraphics[width=0.6\textwidth]{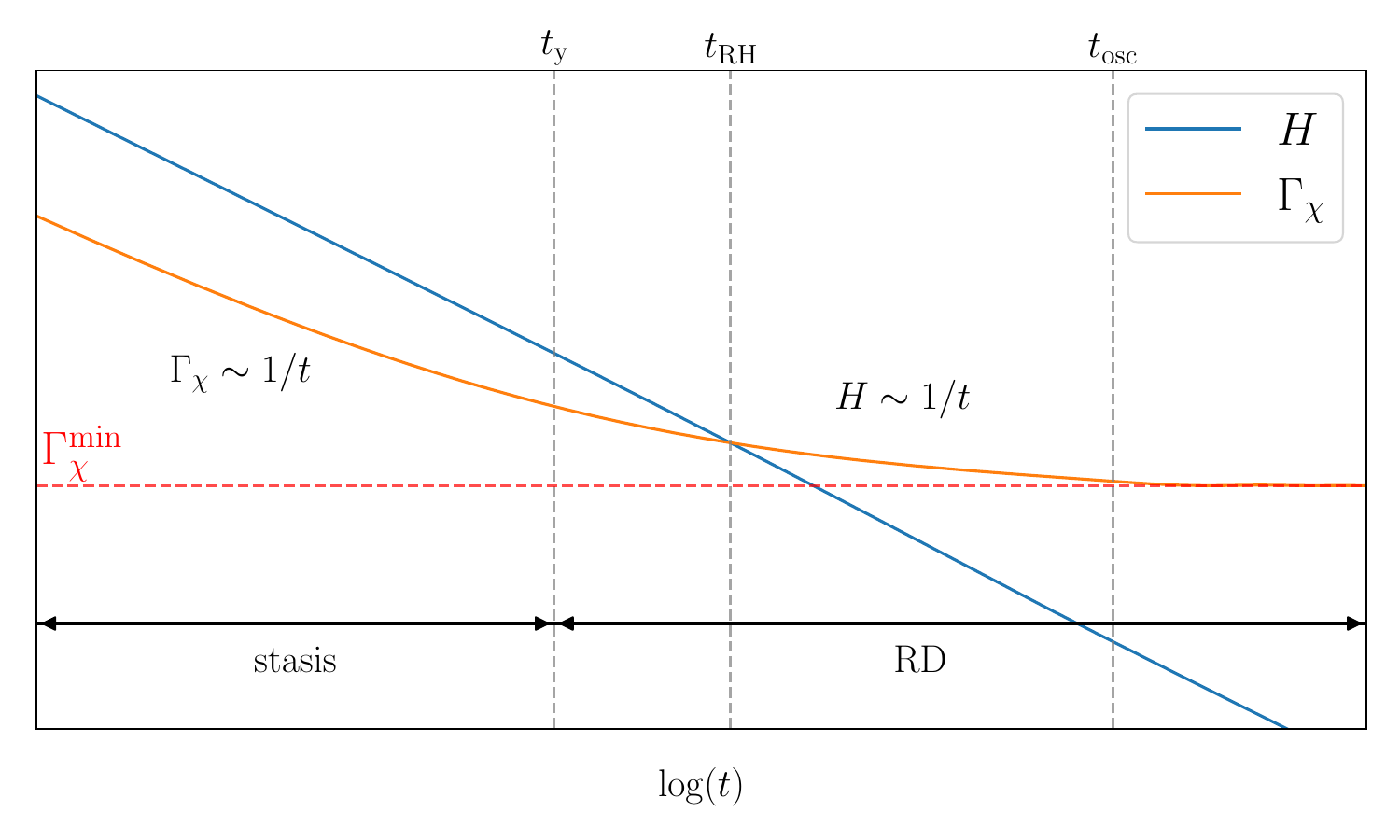}
\includegraphics[width=0.6\textwidth]{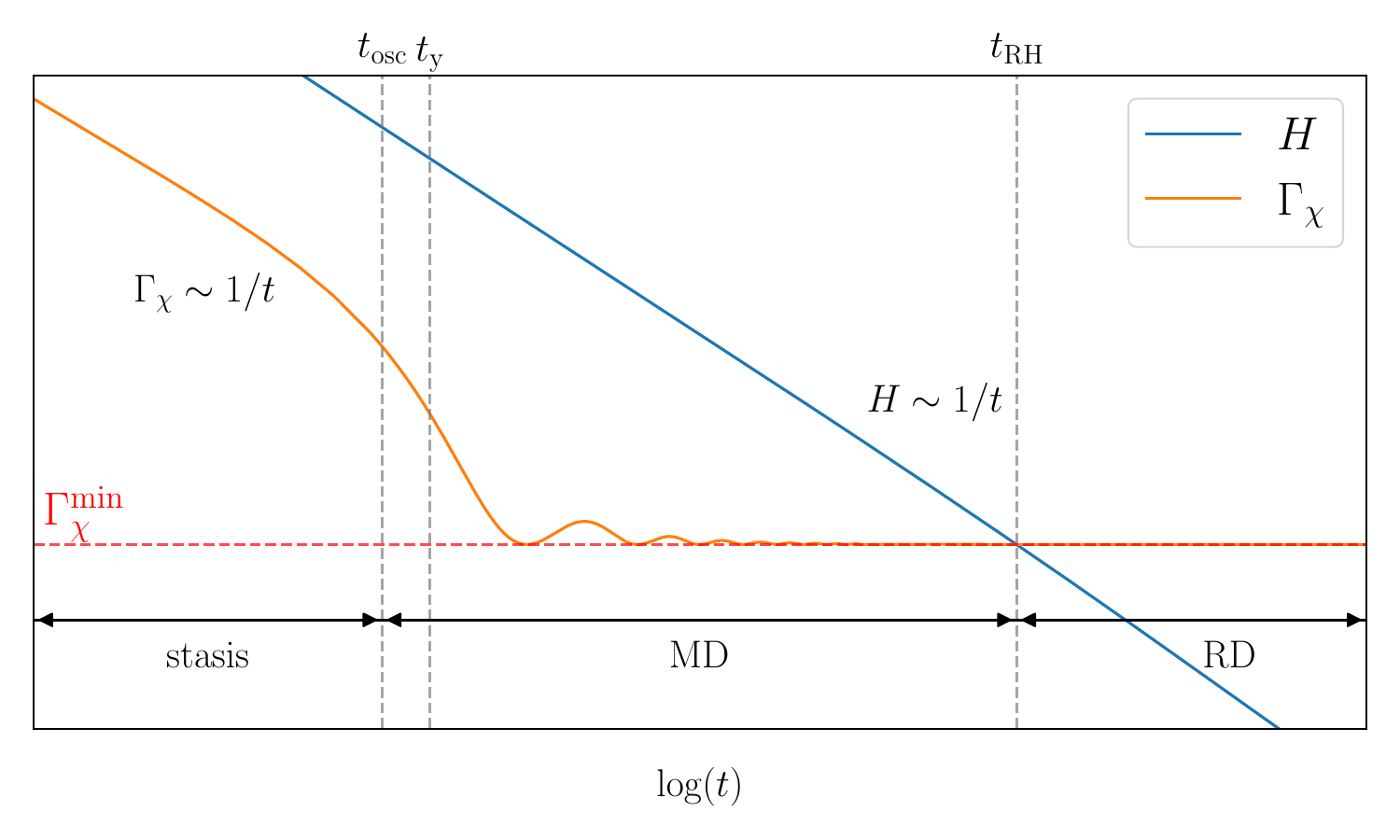}
\caption{Schematic plots of the decay rate $\Gamma_\chi$ and the Hubble rate $H$ as a function of time on logarithmic scale for Case I (RD universe after stasis) and Case II: matter-dominated universe after stasis. 
The red dashed line indicates the minimum decay width $\Gamma_\chi^{\mathrm{min}}$. 
Three critical instants, $t_y$, $t_{\mathrm{osc}}$, and $t_{\mathrm{RH}}$, are indicated by the dashed vertical lines. 
}
\label{fig:ToscTy}
\end{figure}

A sketch of these two cases is shown in Fig.~\ref{fig:ToscTy}. 
In the upper panel, we present Case I, in which stasis ends at $t_y$,
and the universe subsequently enters an RD era at $t_{\rm RH}$ before the scalar field $\phi$ starts oscillating. 
In the lower panel, we present Case II, in which stasis ends at $t_{\rm osc}$, and the universe enters an MD epoch.
Later, at $t_{\mathrm{RH}}$, the rapid decay of $\chi$ reheats the universe into an RD epoch. 
In both cases, the decay width $\Gamma_\chi$ scales like $1/t$ during stasis before it reaches either $t_y$ or $t_{\mathrm{osc}}$.
Note that the reheating time $t_{\rm RH}$ is always later than $t_y$, when the field-dependent coupling ceases to be dominant.
This is because $\Gamma_\chi/H = 1-\overline{\Omega}_\chi<1$ during stasis.
Thus, it takes a while for $\Gamma_\chi$ to reach $H$, and we have $t_y \lesssim t_\mathrm{RH}$ in general.

\begin{figure}[t]
\centering
\includegraphics[width=0.7\textwidth]{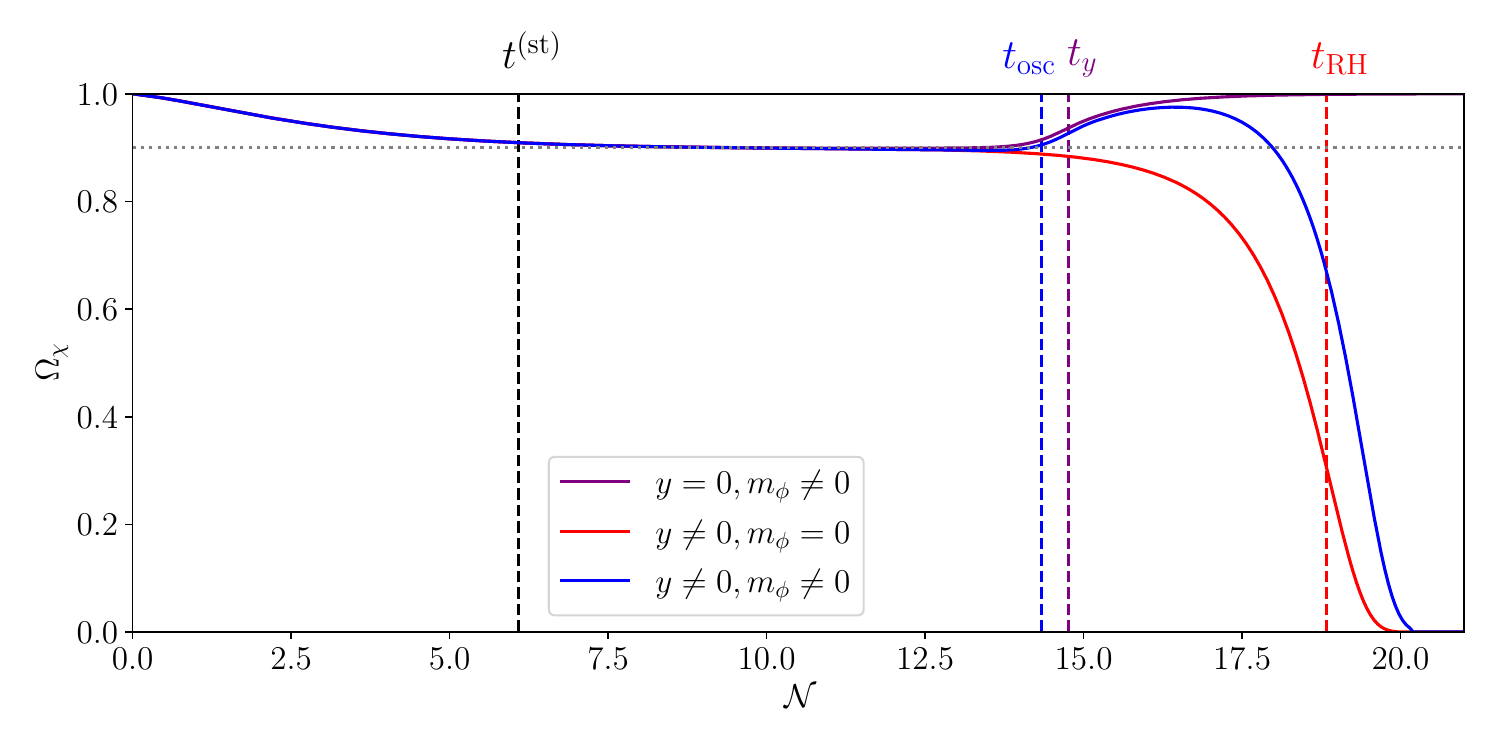}
\caption{Time evolution of $\Omega_\chi$. 
The benchmark parameters are chosen such that $m_\psi = 1\,{\rm GeV}$, $m_\chi = 10^{10}\,{\rm GeV}$, $m_\phi = 100\,{\rm eV}$, $\zeta =1/16$, $g=0.01$, $y = 4 \times 10^{-10}$, $\Lambda_1 = 9.2 \times 10^{18}\,{\rm GeV}$, $\Lambda_2 = 1.5 \times 10^{15}\,{\rm GeV}$, $\Lambda_3 = 10^{16}\,{\rm GeV}$, $H^{(0)} = 250\,{\rm GeV}$, $\rho_\gamma^{(0)} = 0$, $\phi (t^{(0)}) = 10^{14}\,{\rm GeV}$ and $\dot{\phi}(t^{(0)})=0$.
The above parameter choices correspond to $f=1$ and yield a stasis solution with $\overline{\Omega}_\chi= 0.9$ as indicated by the horizontal dotted line.
The three different solid curves represent the different behaviors if we turn on or off $y$ or $m_\phi$. 
The black, blue, purple, and red vertical lines indicate the critical instants 
$t^{(\mathrm{st})}$,
$t_{\mathrm{osc}}$, $t_y$, and $t_\mathrm{RH}$, respectively.
}
\label{fig:Benchmark}
\end{figure}

In Fig.~\ref{fig:Benchmark},
we present the evolution of $\Omega_\chi$ by numerically solving the Boltzmann equations in \eqref{eq:Boltzmann_rhox} and the equation of motion in \eqref{eq:EoM_phi} for the effective potential in Eq.~\eqref{eq:Veff}. 
We have taken into account the temperature dependence in $g_\star(T)$.
In particular, we have used the fitting formula provided in the appendix of Ref.~\cite{Wantz:2009it},
and added to it the contribution from the field $\psi$.
For later convenience, we define the time $t^{(\mathrm{st})}$ at which stasis starts as the instant when $|\Omega_\chi(t)-\overline{\Omega}_\chi|/|\overline{\Omega}_\chi| < 0.01$
is fulfilled for the first time.\footnote{Note that this condition can be fulfilled multiple times if the stasis solution is a spiral point (see Fig.~\ref{fig:stability_scan}).}
The critical moments 
$t^{(\mathrm{st})},\,t_{\mathrm{osc}},\,t_y$ and $t_{\mathrm{RH}}$ are then indicated by the dashed vertical lines.

The solid purple, red, and blue curves stand for cases in which $\{y = 0,m_\phi\neq 0 \}$,
$\{y\neq 0,m_\phi = 0 \}$,
and $\{y\neq 0,m_\phi\neq 0 \}$, respectively,
while the other benchmark parameters are held fixed.
In all cases, the behavior at early times is consistent with the discussion in Sec.~\ref{Sec:StabilityAnalysis} and does not depend on the values of $m_\phi$ or $y$ since the Hubble mass term and the field-dependent coupling are dominant.
However, as discussed above, the values of $\{m_\phi,y\}$ do affect how stasis ends.
For example, as indicated by the purple curve, if $y=0$, the decay width $\Gamma_\chi$ decreases rapidly after stasis ends and
never becomes comparable to the Hubble rate $H$.
The universe, therefore, ends up in an eternal MD epoch.
On the other hand, if $m_\phi=0$, the scalar field $\phi$ never oscillates, 
and, as represented by the red curve, the universe goes to an RD era soon after stasis ends at $t_y$.
In addition, the solid blue curve represents the solution when both $y$ and $m_\phi$ are non-zero. Since in this particular example $t_{\rm osc} < t_y$, the universe starts to enter a matter-dominated epoch after stasis ends, 
and the abundance $\Omega_\chi$ increases. 
However, as the decay width quickly becomes comparable to the Hubble rate at $t_{\mathrm{RH}}$, the universe reheats itself through the decay of $\chi$ and enters an RD era before reaching a complete matter domination. 

Note that since we have assumed that $m_{\psi,\mathrm{th}}^2 = \zeta g^2 T^2 \gg m_\psi^2$ in Eq.~\eqref{eq:Veff}. 
This sets a bound on when our model becomes invalid depending on the values of $g,~\zeta $ and $m_\psi$. 
In particular, for the benchmark values chosen in Fig.~\ref{fig:Benchmark} we need $T\gg 400\,\text{GeV}$ during stasis. 
We have checked that this condition is satisfied for the chosen benchmark values. 
We have also checked that the energy density $\rho_\phi$ does not exceed the matter energy density at the matter-radiation equality (MRE), when the temperature $T_{\mathrm{MRE}} \sim 1$ eV. 
In the following subsections,
we shall study these constraints in more detail for general values of $y$ and $m_\phi$.

\subsection{Duration of stasis}
\label{subsec:DurationStasis}
In both case I and case II, the duration of stasis can be estimated by 
\beq
    \mathcal{N}_s \equiv \log \left( \frac{a_s}{a^{\rm (st)}} \right) \approx \frac{\overline{\kappa} }{3} \log \left( \frac{t_s}{t^{(\mathrm{st})}} \right)\, ,
    \label{eq:Ny}
\eeq
in which $a^{\rm (st)}\equiv a(t^{\rm (st)})$ and $a_s\equiv a(t_s)$, and the end time of stasis
\beq
    t_{s} = \begin{cases}
        t_y,\,\text{ for Case I}\, ,\\
        t_\mathrm{osc},\,\text{ for Case II}\, .
    \end{cases}
    \label{eq:ts}
\eeq
We have also used that $a\sim t^{\overline{\kappa}/3}$ before reaching $t_s$.
Using Eq.~\eqref{eq:tosc} and Eq.~\eqref{eq:ty}, and identify $t_*$ with $t^{\rm (st)}$, 
the number of e-folds during stasis can be rewritten as
\beq
    \mathcal{N}_{s} = \begin{cases}
        \displaystyle\frac{\overline{\kappa}}{3}\log \left( \frac{\phi^{4}(t^{(\mathrm{st})})}{\Lambda_3^4 y^2}\right) & \text{ for Case I}\,,\\
        \displaystyle\frac{\overline{\kappa}}{3}\log\left(\frac{\overline{\kappa}}{2t^{(\mathrm{st})} m_\phi}\right)& \text{ for Case II}\,     .
    \end{cases}
    \label{eq:Ns}
\eeq
Interestingly, in Case I, the term in parentheses is simply the squared ratio of the field-dependent Yukawa coupling $\phi^2/\Lambda_3^2$ at the start of stasis to the minimum Yukawa coupling $y$.
Likewise, in Case II, the ratio in parentheses can be rewritten as $3H(t^{\rm(st)})/(2m_\phi)$, which indicates how much the scalar field would be damped when stasis starts if its mass were solely the classical mass.
It is also proportional to the ratio of the mass of $\phi$ at the start of stasis to its classical mass.
Thus, the duration of stasis $\mathcal{N}_s$ in our scenario reflects the hierarchy between physical quantities at different time or energy scales.

In addition, we are also interested in the temperature of the thermal bath immediately after the universe reheats and transitions into a RD epoch.
Assuming an instantaneous reheating at $t_{\rm RH}$,
the number of e-folds between the end of stasis and reheating is given by
\beqn
\mathcal{N}_{\Delta}&\equiv& \log \left(\frac{a_{\rm RH}}{a_s}\right)\nn\\
&\approx& \begin{cases}
         \displaystyle\frac{\overline{\kappa}}{3}\left[\log(4\pi)+4\log \left( \frac{\Lambda_3 }{  \phi(t^{(\mathrm{st})})}\right)-\log\left(m_\chi t^{(\mathrm{st})}\right)\right] & \text{ for Case I}\,,\\
       \displaystyle\frac{2}{3} \left[ \log(4\pi)+\log \left( \frac{ m_\phi}{ m_\chi} \right)-2\log (y)\right]& \text{ for Case II}\,,
    \end{cases}
    \label{eq:NDelta}
\eeqn
where $a_{\rm RH}\equiv a(t_{\rm RH})$.
Note that, in the above equations, we have made the approximation that, during the transition epoch from the end of stasis to the reheating time, $\kappa\approx \overline{\kappa}$ for Case I, and $\kappa=2$ for Case II.

The reheating temperature $T_{\rm RH}$ of the universe can be obtained from the energy density of the thermal bath. 
Using the approximation that the reheating is instantaneous,
we have
\beq
\rho(t_\mathrm{RH}^{+}) =\rho_\gamma (t_\mathrm{RH}^{+})=\rho (t_\mathrm{RH}^{-}) = \rho_\gamma (t_\mathrm{RH}^{-})+\rho_\chi (t_\mathrm{RH}^{-})\, ,
    \label{eq:rhotRD0}
\eeq
where the superscript ``$-$'' or ``$+$'' denotes the instant immediately before or after the reheating time $t_{\rm RH}$.
Together with Eqs.~\eqref{eq:Ns} and \eqref{eq:NDelta},
the energy density of the universe at reheating can be estimated from the matter and radiation energy densities at $t^{(\mathrm{st})}$ by scaling down to $t_\mathrm{RH}^{-}$, which leads to
\beqn
\rho(t_\mathrm{RH}^+)=\rho(t_\mathrm{RH}^-)
&\approx& \rho_{\gamma}(t^{(\mathrm{st})})\bigg(\frac{a^{(\rm st)}}{a_s}\bigg)^{-3(1+\overline{w})}\left(\frac{a_{\rm RH}}{a_s}\right)^{-4}+
\rho_{\chi}(t^{(\mathrm{st})})\bigg(\frac{a^{(\rm st)}}{a_s}\bigg)^{-3(1+\overline{w})}\left(\frac{a_{\rm RH}}{a_s}\right)^{-3}\nn\\
&\approx&\rho_{\gamma}(t^{(\mathrm{st})})e^{-3 (1+\overline{w})\mathcal{N}_s}e^{-4\mathcal{N}_\Delta}+\rho_\chi(t^{(\mathrm{st})})\mathrm{e}^{-3 (1+ \overline{w})\mathcal{N}_s}e^{-3\mathcal{N}_\Delta}\, .\label{eq:rhotRD}
\eeqn
The temperature $T_{\rm RH}$ can then be obtained from above using $\rho(t_\mathrm{RH}^+)=\frac{\pi^2}{30}g_\star(T_\mathrm{RH}) T_\mathrm{RH}^4$.

\subsection{Constraints from the relic abundance of \texorpdfstring{$\phi$}{phi}}
\label{subsec:BoundrhoPhi}
Although the energy density of the scalar field $\phi$ is assumed to be negligible during stasis, its relative abundance can, in general, grow during stasis and even after reheating.
It is, therefore, important to ensure that the relic abundance of $\phi$ is consistent with observational data on the expansion history and the present-day energy content of our universe.

The population of $\phi$ can be separated into two parts --- the zeroth-mode field which rolls down the Hubble mass potential and controls the decay widths of $\chi$ during stasis, and the nonzero momentum $\phi$ particles produced by the decay of $\chi$.
In what follows, we shall consider them separately.

\subsubsection*{Zero-momentum mode}
As discussed in Sec.~\ref{sec:field_dependent_decay}, the zero-momentum mode of $\phi\sim t^{-1/4}$ during stasis.
After that, the universe may enter directly an RD epoch,
or an MD epoch before reheating, depending on how stasis ends.
In any case, since the scalar field $\phi$ eventually oscillates and behaves like non-relativistic matter,
it can be considered as a potential component of dark matter.
Therefore, we can obtain a constraint on the relic abundance of the zero-momentum mode $\phi$ by evaluating its energy density at MRE to ensure that $\Omega_\phi(t_{\rm MRE})\leq \Omega_{\rm DM}(t_{\rm MRE})\approx 1/2$.

The energy density of $\phi$ by the end of stasis can be estimated straightforwardly by scaling down the field value $\phi(t^{\rm (st)})$ at the beginning of stasis.
However, its subsequent evolution after stasis ends depends on the order of the critical moments $t_{\rm osc},~t_y,~t_{\rm RH}$, and $t_{\rm MRE}$.
To address the evolution of $\phi$ in various cases,
let us first generalize the result in Eq.~\eqref{eq:general_phi}.
In particular, since we have not made any assumption about the value of $\overline{\kappa}$, it is easy to see that Eq.~\eqref{eq:general_phi} applies to any background cosmology described by $H=\kappa/(3t)$ with an arbitrary constant $\kappa$.
Consequently, we find that not only $\phi\sim t^{-1/4}$ during stasis,
the field $\phi$ follows the same scaling relation in time even in the RD or MD epoch as long as the Hubble mass term dominates its potential.
We can, therefore, parametrize the field evolution during any epoch characterized by a constant $\kappa$ as
\beq
\phi = \phi(t_{\rm ref}) \bigg( \frac{t_{\rm ref}}{t}\bigg)^{1/4}\, ,
\label{eq:phiDuringStasis}
\eeq
where $t_{\rm ref}$ is an arbitrary reference time within the epoch in consideration.
The kinetic and potential energies of $\phi$ are then given by
\beqn
K(\phi) &\equiv& \frac{1}{2} \dot{\phi}^2 =  \frac{\phi^2 (t_{\rm ref})}{32t^2}\bigg(\frac{t_{\rm ref}}{t}\bigg)^{1/2}\, ,
\label{eq:KineticEnergy}\\
V(\phi) &=& \lambda H^2 \phi^2(t) = 
\lambda \frac{\kappa^2 \phi^{2} (t_{\rm ref})}{9 t^{2}}\left(\frac{ t_{\rm ref}}{t}\right)^{1/2}\, .
\label{eq:PotentialEnergy}
\eeqn
Apparently, the kinetic energy and potential energy of $\phi$ are proportional to each other. 
The equation-of-state parameter of the scalar field $\phi$ is, therefore, a constant given by
\beqn
w_\phi &=& \frac{K(\phi)-V(\phi)}{K(\phi)+V(\phi)} = \frac{9 - 32 \lambda \kappa^2 }{9 + 32 \lambda \kappa^2} \, .
\label{eq:wphiconst}
\eeqn
Using Eq.~\eqref{eq:kappa_x} with $\lambda_t=\overline{\kappa}^2\lambda/9$ and $x=2$,
we can further simplify the above equation and find that 
\beq
w_\phi=-\frac{2\kappa-3}{2\kappa-2}=-\frac{3\Omega_\chi}{4+2\Omega_\chi}\leq 0\,,
\eeq
where, in the last step, we have used the relation between $\kappa$ and $\Omega_\chi$ in Eq.~\eqref{eq:kappa_Omegachi}.
The above equation suggests that $w_\phi$ is independent of $\lambda$,
but depends on the composition of the universe instead as a decreasing function of $\kappa$ (or $\Omega_\chi$).  
In addition, comparing with the equation-of-state parameter of the background universe, which is given by $w=2/\kappa -1$,
it is easy to prove that $w_\phi<w$ for $\kappa>4/3$.
Such condition is always satisfied for a universe dominated by an arbitrary fixed mixture of matter and radiation.
As a result, $\Omega_\phi$ always grows with time, no matter whether the background universe is radiation-dominated, matter-dominated, or in stasis.

We are now equipped to evaluate the energy density $\phi$ at MRE.
There can be four different cases.
\begin{enumerate}
    \item $t_\mathrm{osc} < t_y < t_\mathrm{RH} < t_\mathrm{MRE}$: In this case, stasis ends at $t_\mathrm{osc}$, and then $\phi$ becomes matter-like with $\rho_\phi \sim a^{-3}$. 
    We therefore have 
    \beq
    \rho_{\phi}(t_\mathrm{MRE}) \simeq \rho_\phi (t_\mathrm{osc}) e^{-3 \mathcal{N}_\Delta}\left(\frac{T_\mathrm{MRE}}{T_\mathrm{RH}}\right)^3 \frac{g_{\star,s}(T_\mathrm{MRE})}{g_{\star,s}(T_\mathrm{RH})} \, ,
    \label{eq:RhoPhiCaseA}
    \eeq
    where $g_{\star,s}$ is the effective number of relativistic freedom for entropy density, and we have used entropy conservation, $g_{\star,s}(T)T^3 a^3=const.$, after reheating.
    \item $t_y  < t_\mathrm{osc} < t_\mathrm{RH} < t_\mathrm{MRE}$: In this case, stasis ends at $t_y$, and $w_\phi$ evolves towards a value larger than $\overline{w}_\phi$ when $t_y <t<t_{\rm osc}$ since $\Omega_\chi$ decrease as $\Gamma_\chi$ reaches its minimum.
    For a conservative bound, we can make the approximation between $t_y$ and $t_{\rm osc}$ that $w_\phi=\overline{w}_\phi$ which underestimates the dilution of $\rho_\phi$. 
    We, therefore, obtain that
    \beq
    \rho_{\phi}(t_\mathrm{MRE}) \lesssim \rho_\phi (t_y ) e^{-3 (1 + \overline{w}_\phi)\mathcal{N}_{y,\mathrm{osc}}} e^{-3 \mathcal{N}_{\mathrm{osc, RH}}}\left(\frac{T_\mathrm{MRE}}{T_\mathrm{RH}}\right)^3 \frac{g_{\star,s}(T_\mathrm{MRE})}{g_{\star,s}(T_\mathrm{RH})}\,,
    \label{eq:RhoPhiCaseB}
    \eeq
    in which we introduce the notation that $\mathcal{N}_{i,j}$ is the number of $e$-folds between two critical instants $t_i$ and $t_j$.
    \item $t_y  < t_\mathrm{RH} <  t_\mathrm{osc} < t_\mathrm{MRE}$: In this case, stasis also ends at $t_y$, but the oscillation of $\phi$ occurs after reheating. Like the previous case, we can set $w_\phi=\overline{w}_\phi$ between $t_y$ and $t_{\rm osc}$ and obtain that
    \beq
    \rho_{\phi}(t_\mathrm{MRE}) \lesssim \rho_\phi (t_y) e^{-3 (1 + \overline{w}_\phi) (\mathcal{N}_\Delta+ \mathcal{N}_{\rm RH, osc})}\left(\frac{T_\mathrm{MRE}}{T_\mathrm{osc}}\right)^3 \frac{g_{\star,s}(T_\mathrm{MRE})}{g_{\star,s}(T_\mathrm{osc})}\, .
    \label{eq:RhoPhiCaseC}
    \eeq
    \item $t_y < t_\mathrm{RH} < t_\mathrm{MRE}  < t_\mathrm{osc}$:
    In this case, $\phi$ continues to slow roll even after the $t_{\rm MRE}$.
    Such a case only happens when the potential of $\phi$ continues to be dominated by the Hubble mass term after MRE.
    However, since the Hubble mass term originates from the coupling of $\phi$ to the energy densities of different components,
    whereas $\chi$ has already decayed after $t_{\rm RH}$,
    the only source that can contribute to the Hubble mass term after reheating is the coupling to the thermal bath through $\psi$.
    On the other hand, the validity condition for the Hubble mass term,
    \ie $m_{\psi,\mathrm{th}}^2 = \zeta g^2 T^2 \gg m_\psi^2$, 
    suggests that $\psi$ needs to stay relativistic and in thermal equilibrium with the SM thermal bath after $t_{\rm MRE}$.
    Such a situation will severely violates existing bounds on the effective number of neutrino species $N_{\rm eff}$ (see, \eg Ref.~\cite{Planck:2018vyg}).
    We shall, therefore, exclude this case.
    Note that, since $t_{\rm osc}\sim 1/m_\phi$, while $T_{\rm MRE}\sim 1~{\rm eV}$, the constraint which ensures that this case will never occur can be roughly given by $m_\phi\gtrsim 10^{-28}~{\rm eV}$.
\end{enumerate}

\subsubsection*{Nonzero-momentum modes}
$\phi$ particles with nonzero momenta are produced through the decay channels of $\chi$ described in Eqs.~\eqref{eq:4-body} and~\eqref{eq:2-bodyClosedFermion}. 
Therefore, in addition to the $\psi$ particles, some fraction of the energy density of $\chi$ is converted to $\phi$ particles with nonzero momentum at reheating. 
Nevertheless, since $m_\phi\ll m_\chi$, these $\phi$ particles are produced relativistically with a typical momentum $p_\phi\sim \mathcal{O}(m_\chi)$.
Let us denote these $\phi$ particles $\phi_k$.
Therefore, the fraction of $\phi_k$ in the total energy density of radiation immediately after the reheating is given by 
\begin{equation}
\label{eq:rhoFracKneq0}
    \frac{\rho_{\phi_k}}{\rho_\gamma}\bigg|_{t_\mathrm{RH}^+} = \frac{1}{2}\frac{\mathrm{Br}\left(\chi\to\bar\psi\psi\phi\phi\right)+ 2\mathrm{Br}\left(\chi\to\phi\phi\right)}{\mathrm{Br}\left(\chi\to\bar\psi\psi\phi\phi\right)+\mathrm{Br}\left(\chi\to\phi\phi\right)+\mathrm{Br}\left(\chi\to\bar\psi\psi\right)}\,,
\end{equation}
where $\mathrm{Br}(i\to f)$ denotes the branching ratio of the process $i\to f$. 
Notice that, this fraction remains fixed up to changes in $g_\star$ as long as $\phi_k$ is relativistic.
After $\phi_k$ becomes non-relativistic, the ratio between $\rho_{\phi_k}$ and the energy density of the matter component will also remain essentially fixed.
Therefore, as long as relativistic/non-relativistic transition occurs later than MRE, the final contribution of $\rho_{\phi_k}$ to the energy density of matter can never exceed the fraction in Eq.~\eqref{eq:rhoFracKneq0}.
Using that $p_\phi \sim a^{-1}\sim T$, the critical temperature $T_\mathrm{NR}$ at  which $\phi_k$ becomes non-relativistic can be estimated by
\begin{equation}
    \label{eq:TnonRel}
    T_\mathrm{NR} \sim T_{\mathrm{RH}}\times \mathcal{O}\left(\frac{m_\phi}{m_\chi}\right)\,.
\end{equation}
We have checked that, for the values considered in this work, we always obtain $T_{\rm NR}<T_{\rm MRE}$.
Thus, this transition always occurs after MRE, which suggests that $\phi_k$ contributes to the radiation component universe as a species of dark radiation.
Considering that the bounds on the extra effective number of neutrino species $\Delta N_{\rm eff}\equiv N_{\rm eff}-3.046$ is $\mathcal{O}(0.1)$ \cite{Planck:2018vyg}, and that $g_\star(T)\lesssim\mathcal{O}(10)$ during BBN, 
a conservative bound on the relic abundance of $\phi_k$ can then be placed by requiring that the fraction
$\left.\rho_{\phi_k}/\rho_\gamma\right|_{t_\mathrm{RH}^+}< 1\%$.

\subsection{Summary of constraints}\label{subsec:Summary_Constraints}

\begin{figure}[t]
\centering
\includegraphics[width=0.48\textwidth]{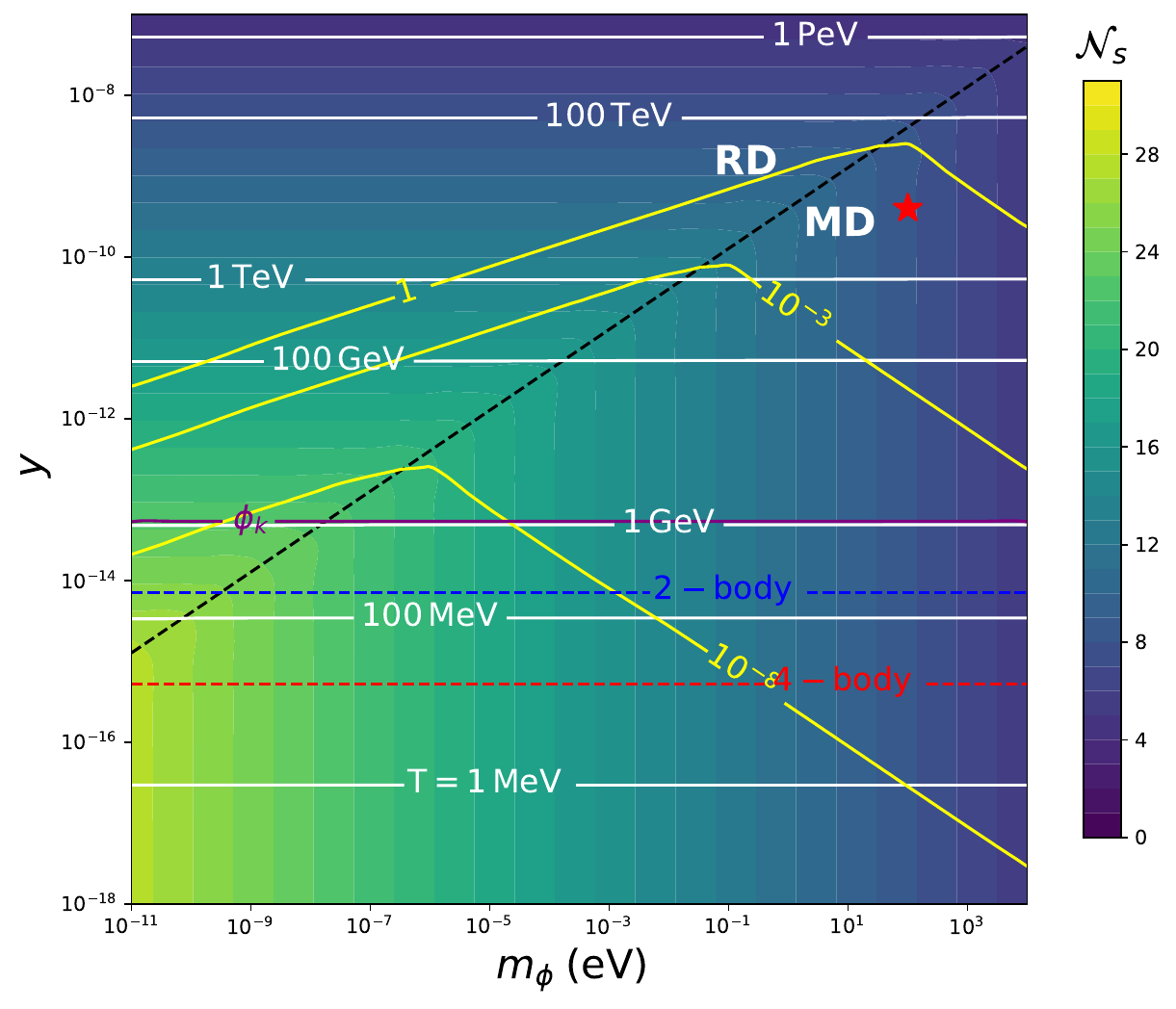}
\includegraphics[width=0.48\textwidth]{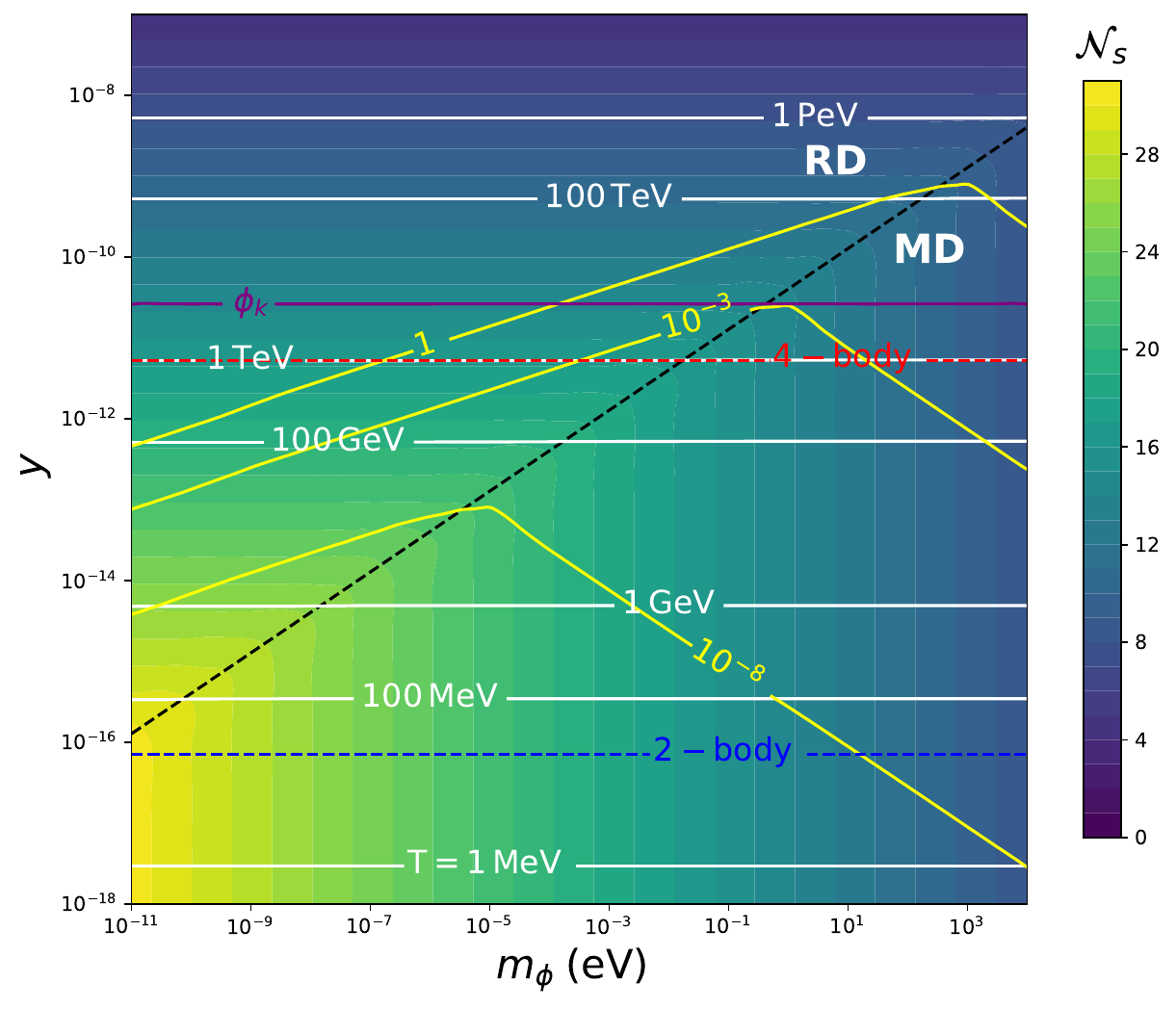}
\caption{Parameter scan in the $m_\phi$-$y$ plane.
In both panels, the color coding indicates the duration of stasis $\mathcal{N}_s$,
while the dashed black line separates the region for which the universe enters an RD or MD epoch after stasis ends.
The white contours represent the reheating temperature $T_{\rm RH}$, and the yellow contour indicates the fraction of dark matter in the form of the oscillating zero-momentum mode of $\phi$.
Moreover, the blue and red dashed lines correspond to the values of $y$ below which $\Gamma_{\chi\to\bar\psi\psi\phi\phi}$ and $\Gamma_{\chi\to\phi\phi}$ become larger than $\Gamma_{\chi\to\bar\psi\psi}^{\rm min}$.
Finally, the purple line represents the value of $y$ below which the fraction of energy density in radiation carried by $\phi$ particles with non-zero momenta at reheating, \ie  
$\left.{\rho_{\phi_k}}/{\rho_\gamma}\right|_{t_\mathrm{RH}^+}$, exceeds $1\%$.
In the left panel,
the values of the parameters are the same as the ones chosen in Fig.~\ref{fig:Benchmark} except for $m_\phi$ and $y$, which we allow to vary.
The benchmark in Fig.~\ref{fig:Benchmark} is represented by the red star.
The right panel is similar to the left panel except that $m_\chi =10^{12}$ GeV and $H^{(0)} = 25 \times 10^{3}$ GeV. In both panels, we have $\overline{\Omega}_\chi=0.9$. }
\label{fig:Nend}
\end{figure}

We present in Fig.~\ref{fig:Nend} two parameter scans on the plane spanned by $m_\phi$ and $y$, which control how stasis ends.
In the left panel, all parameters except $m_\psi$, $m_\phi$ and $y$ are set to the values used in Fig.~\ref{fig:Benchmark}, which also correspond to $\overline{\Omega}_\chi=0.9$ with the stasis solution being a stable node.
The benchmark in Fig.~\ref{fig:Benchmark} is marked by the red star.
The right panel is similar, but we choose a different mass $m_\chi$ and a different initial Hubble rate $H^{(0)}$.
Note that this change does not affect $\overline{\Omega}_\chi$.
In both panels, we do not specify the value of $m_\psi$ since the dynamical evolution of the system is not affected by it as long as $\psi$ stays relativistic.
The consistency of our scenario can be ensured by choosing a sufficiently small $m_\psi$ that satisfies $m_{\psi,\rm th}> m_\psi$ at reheating.
In both panels, we indicate the duration of stasis $\mathcal{N}_s$ by the color coding, while the reheating temperature $T_{\rm RH}$ and the fraction of dark matter in the oscillating zero-momentum mode of $\phi$ are shown by the white and yellow contours, respectively.
Notice that, since reheating happens when $H\sim \Gamma_\chi$, where $H$ encodes the energy density and $\Gamma_\chi$ is dominated by the Yukawa coupling $y$, the reheating temperature mainly depends on the value of $y$ and is insensitive to $m_\phi$. 
In addition, depending on whether the Yukawa coupling $y$ or the scalar mass $m_\phi$ triggers the end of stasis first, the region of parameter space is divided diagonally by the dashed black line.
Indeed, as discussed in Sec.~\ref{sec:exit_stasis}, the upper left and lower right regions correspond to Case I (RD epoch after stasis) and Case II (MD epoch after stasis), respectively.
In the upper left region, the fraction of dark matter in the zero-momentum mode of $\phi$ decreases as one increases $m_\phi$ while holding $y$ fixed.
This is reasonable since a larger $m_\phi$ corresponds to an earlier oscillation time which leads to a stronger dilution effect.
On the other hand, the behavior in the lower right region is different --- the fraction of dark matter in $\phi$ increases with $y$ if $m_\phi$ is held fixed.
This is because a larger $y$ corresponds to an earlier reheating. And since $\phi$ has already started oscillating and become matter-like before reheating, its abundance will increase more relative to that of radiation if reheating starts earlier.

Additionally, we also check the consistency of our scenario by considering competing decay channels of $\chi$. 
In particular, both $\Gamma_{\chi\to\bar\psi\psi\phi\phi}$ and $\Gamma_{\chi\to\phi\phi}$ must be subdominant compared to $\Gamma_{\chi\to\bar\psi\psi}$.
As a conservative bound, dashed red and blue lines indicate the values of $y$ below which the competing 4-body and 2-body decay widths become larger than the minimum of the main decay width $\Gamma_{\chi\to\bar\psi\psi}^{\rm min}$.
Notice that, $\Gamma_{\chi\to\bar\psi\psi}^{\rm min}$ depends only on $y$ instead of the field-dependent coupling $(\phi/\Lambda_3)^2$.
Therefore, although violating this conservative bound would shorten the duration of stasis, it does not generally prevent the occurrence of the stasis phenomenon.
Moreover, as both competing channels produce $\phi$ particles with non-zero momenta, we compute the fraction of radiation energy density in $\phi_k$ after reheating using Eq.~\eqref{eq:rhoFracKneq0}.
The purple contour corresponds to a reference below which the ratio $\left.{\rho_{\phi_k}}/{\rho_\gamma}\right|_{t_\mathrm{RH}^+}$ exceeds $1\%$.
Notice that this contour is nearly horizontal since the Yukawa coupling typically dominates the decay width of $\chi$ at reheating.

Taken together, the viable parameter space for our stasis scenario, therefore, corresponds to the region below the yellow contour for which $\rho_\phi(t_\mathrm{MRE})/ \rho_\gamma(t_\mathrm{MRE}) < 1$ and above the red, blue, and purple lines. 
Since we choose not to specify the value of $m_\psi$,
the condition that $m_{\psi,\rm th}> m_\psi$ at reheating translates into a condition that $T_{\rm RH}> 400 m_\psi$.
Such condition is clearly satisfied by the benchmark in Fig.~\ref{fig:Benchmark} with $m_\psi= 1 {\rm GeV}$ as the reheating temperature associated with the location marked by the red star is well above $\mathcal{O}({\rm TeV})$.
Even for the smallest reheating temperature $T_{\rm RH}\sim 1 {\rm GeV}$ in the allowed region in the left panel, a mass $m_\psi\sim \mathcal{O}({\rm MeV})$ can satisfy the constraint.
We see that the reheating temperature and the constraints change for different values of $m_\chi$ or $H^{(0)}$. 
Furthermore, we point out that if we were to choose a smaller value of $\overline{\Omega}_\chi$, the qualitative behavior of Fig.~\ref{fig:Benchmark} would not change, but the duration of stasis would be smaller for the same values of $y$ and $m_\phi$ since it would take longer to reach stasis.

Overall, we see that stasis can easily last 10 to 20 e-folds.
We also note that if the reheating temperature is smaller than the QCD crossover temperature around $200$ MeV, $f$ may experience a drastic change since it depends on $g_\star(T)$ (see Eq.~\eqref{eq:para_identification}). 
In this case, $f$ will increase which leads to a higher value of $\overline{\Omega}_\chi$ (see Fig.~\ref{fig:stability_scan}).
Furthermore, the energy density of $\psi$ should be sufficiently small before BBN starts around $1$ MeV. 
The allowed parameter space for the mass $m_\psi$ and the coupling $g$ to the thermal bath depend heavily on the way in which $\psi$ couples to the thermal bath. 
For example,
if $\psi$ is a dark-matter component, its mass and coupling can be constrained by direct and indirect detection experiments, and collider searches (see, \eg Ref.~\cite{Cooley:2022ufh} and references therein).
It would be interesting to constrain further the values in our model by specifying how $\psi$ couples to the thermal bath, but these questions go beyond the scope of this work.

\FloatBarrier

\section{Conclusion}\label{sec:conclusion}

Cosmological stasis has been shown to be a possible prediction of many BSM models within a wide range of parameter space.
In this paper, we have demonstrated that a stasis between matter and radiation in the early universe can also arise when the decay of the matter component is regulated by the dynamics of a scalar field $\phi$ with negligible abundance.
Unlike previous literature, such a stasis does not require the existence of a large tower with an extended mass spectrum, 
nor does it rely on the thermalization of the matter fields.
Instead, the matter/radiation stasis in our work arises due to the interplay between the field-dependent coupling, which controls the decay width, and the potential of $\phi$, which depends on the Hubble rate.
Through stability analysis, we have found that, depending on the value of $\overline{\Omega}_\chi$, the associated fixed point is either a stable global attractor or an unstable solution that exhibits undamped oscillatory behavior. 
In the stable regime, we showed that the system can approach stasis either asymptotically or exhibit intrinsic underdamped oscillations.
Such properties are quite distinct from all previous stasis scenarios in which stasis is typically a stable node for the entire physical range $0<\overline{\Omega}_\chi<1$,
and the oscillatory behavior observed in the tower-based stasis scenarios \cite{Dienes:2023ziv} can only appear when the mass spectrum is no longer effectively continuous for relevant timescales.

We have shown that this realization of matter/radiation stasis can arise naturally in models motivated by supergravity.
We have presented a minimal model which consists of a heavy real scalar $\chi$, a Dirac fermion $\psi$, and a light real scalar $\phi$.
In this model, the heavy matter field $\chi$ converts its energy into radiation by decaying primarily into a pair of $\psi$ particles, which are assumed to be in thermal equilibrium with the thermal bath. 
The decay width of $\chi$ is mediated by a field-dependent coupling that depends on the field value of the zero-momentum mode of the light scalar $\phi$, 
where $\phi$ is effectively coupled to the energy densities of $\chi$ and $\psi$ such that it acquires an effective Hubble mass. 
Furthermore, the model also provides two ways to exit stasis, which is necessary for a viable cosmological scenario.
On the one hand, stasis can end if the decay width $\Gamma_\chi$ is no longer dominated by the field-dependent coupling but by a direct Yukawa coupling $y$.
In this case, $\Gamma_\chi$ will stop decreasing,
and the universe will end up in an RD epoch.
On the other hand, stasis can also end if the Hubble mass becomes subdominant compared to the classical mass $m_\phi$.
In this second case, the scalar field will start underdamped oscillation, which quickly minimizes the decay width $\Gamma_\chi$ such that the universe enters a matter-dominated epoch as the abundance of $\chi$ keeps increasing.
In either case, the universe will be reheated and enter the RD epoch due to the direct coupling $y$.

In this paper, we have also discussed various constraints on our stasis model.
In particular, to avoid spoiling BBN, 
the reheating temperature by the end of stasis, or by the end of the MD epoch after stasis has to be above $\mathcal{O}({\rm MeV})$.
Besides the reheating temperature, the relic abundance of $\phi$ also provides another set of constraints.
In one aspect, the abundance of the zero-momentum mode of $\phi$ always grows relative to the other abundances,
and, once it starts oscillating, it can be a (subdominant) component of dark matter.
Therefore, not only do we demand that its abundance be negligible during stasis, but we also require that its energy density be smaller than that of radiation at MRE.
In another aspect, $\phi$ particles with large momenta (denoted as $\phi_k$) are also produced via the decay of $\chi$.
These relativistic non-interacting particles contribute to the radiation abundance as dark radiation.
For a conservative bound, we require that the energy density of $\phi_k$ is less than $1\%$ in that of the entire radiation component immediately after reheating.

Aside from the constraints needed for joining the standard $\Lambda$CDM timeline,
we have also analyzed the conditions required for the self-consistency of our stasis scenario.
One key ingredient of our stasis mechanism is that the decay width $\Gamma_\chi$ is dominated by the field-dependent decay channel $\chi\to\bar\psi\psi$ for which the field-dependent coupling $(\phi/\Lambda_3)^2$ must also be dominant.
In addition, scattering processes involving $\chi$ and $\phi$ must also be suppressed to avoid thermalization.
Another crucial ingredient is the validity of the Hubble mass term in the scalar potential $V(\phi)$.
This validity is ensured if the thermal mass $m_{\psi,{\rm th}}$ dominates over the classical mass $m_\psi$ during stasis, which also ensures that the $\psi$ particles are relativistic.
As we have seen in Fig.~\ref{fig:Nend}, 
all the phenomenological and model-consistency constraints can be satisfied in the viable region of the parameter space.

Additionally, the mass of $\psi$ and its coupling to the thermal bath $g$ can be further constrained depending on the way it is coupled to the SM fields.
For instance, if $\psi$ is a component of dark matter, it can be constrained by direct-/indirect-detection experiments.
These intriguing questions have been studied in various contexts independent of cosmological stasis, and are beyond the scope of this paper.

The existence of such a stasis epoch may lead to interesting phenomenological implications,
many of which have already been mentioned in Ref.~\cite{Dienes:2021woi,Dienes:2022zgd,Dienes:2023ziv,Dienes:2024wnu,Batell:2024dsi}.
In our explicit model, since $\phi$ is a potential dark-matter candidate,
its spectrum of density perturbations today may contain imprints of the earlier non-standard expansion history.
It is, therefore, of great interest to study how the growth of different perturbation modes are affected by the stasis epoch
and whether or not the perturbation spectrum in our scenario may contain distinguishable features that differ from those studied in other stasis scenarios \cite{Dienes:2025tox}.

\acknowledgments
We thank Michael Ratz for initial collaboration and insightful discussions.
We also thank Keith R. Dienes, Max Fieg, Minyuan Jiang, and Brooks Thomas for fruitful discussions.
FH would also like to acknowledge the hospitality of the Center for High Energy Physics, Peking University.
The research activities of FH are supported by ISF Grant 1784/20, MINERVA Grant 714123, and MINERVA grant, project 7141230301.
The work of V.K.-P. was supported by the U.S.\ National Science Foundation under Grant No.\ PHY-2210283 and by UC-MEXUS-CONACyT grant No. CN-20-38.

\bibliographystyle{JHEP}
\bibliography{main}

\end{document}